\documentclass{aa}

\usepackage{graphicx}
\usepackage{supertabular}
\usepackage{natbib}
\newcommand {\hI} {H\,I\,\,} 
\newcommand {\hII} {H\,II\,\,}  
\newcommand {\ha} {H$\alpha$\,\,}  
\newcommand {\kms} {\,km\,s$^{-1}$\,}  
\newcommand {\M} {\mbox{${\cal M}$}}  
  
\newcommand {\msol} {\M$_\odot$\,}  
\newcommand {\lsol} {\L$_\odot$\,}

\begin{document}  

\title{Accurate Determination of the Mass Distribution in Spiral Galaxies:}

\subtitle{3. Fabry-Perot Imaging Spectroscopy of 6 Spiral Galaxies}  
   
\author{S. Blais--Ouellette\inst{1}\inst{2}\inst{3} \and P. Amram\inst{2} \and
C. Carignan \inst{3} \and R. Swaters \inst{4}\thanks{SBO, PA and CC,
Visiting Astronomers, Canada--France--Hawaii Telescope, operated by
the National Research Council of Canada, the Centre National de la
Recherche Scientifique de France, and the University of Hawaii.}}

\offprints{S. Blais--Ouellette}

\institute{Department of Astronomy, California Institute of
  Technologies, CA 91125, USA
\and Laboratoire d'Astrophysique de Marseille, 2 Place Le Verrier,
F--13248 Marseille Cedex 04, France
\and D\'epartement de physique and Observatoire du mont M\'egantic,
Universit\'e de Montr\'eal, C.P. 6128, Succ. centre ville, Montr\'eal,
Qu\'ebec, Canada. H3C 3J7
\and Carnegie Institution of Washington, Washington DC 20015, USA}
   
\date{Received / Accepted}

\renewcommand{\thefootnote}{\alph{footnote}}

\maketitle

\abstract{ High resolution Fabry--Perot data of six spiral galaxies
  are presented. Those data extend the previous sample of spiral
  galaxies studied with high resolution 3--D spectroscopy to earlier
  morphological types. All the galaxies in the sample have available
  \hI data at 21 cm from the VLA or Westerbork. Velocity fields are
  analyzed and \ha rotation curves are computed and compared to \hI
  curves. The kinematics of NGC 5055 central regions are looked at
  more closely. Its peculiar kinematics can be interpreted either as a
  bipolar outflow or as a counter-rotating disk, possibly hosting a
  $9\pm2 \times 10^8$\msol compact object.
  
  Most of the \ha rotation curves present a significantly steeper
  inner slope than their \hI counterparts. The 21 cm data thus seems
  affected by moderate to strong beam smearing. The beam smearing has
  an effect at higher scale-length/beam-width than previously thought
  (up to 20 km/s at a ratio of 8.5).

\keywords{galaxies: fundamental parameters (masses) --
        galaxies: individual: IC 2574, NGC 3109,
        UGC 2259, NGC 5585, NGC 6503, NGC 2403, NGC 6946, NGC 3198,
        NGC 5055, NGC 2841, NGC 5985 --
        galaxies: kinematics and dynamics --
        cosmology: dark matter --
        techniques: radial velocities
        }
}

\section{Introduction}

$\Lambda$CDM N-body numerical simulations, predict that the innermost
parts of density profiles are cuspy: e.g. \cite{NFW96} predict an
inner slope $\alpha=-1.0$ while \cite{Moore99} finds an inner slope
$\alpha=-1.5$. On the other hand, the observational results do not
confirm these cosmological predictions but strongly suggest that dark
halos have a constant central density \cite[inner slope closer to
$\alpha=0$ than to $\alpha=-1$; e.g.][]{deBlok03,
  Swaters03,paperI,paperII}.  Despite the observational evidences that
the central density profiles of dark matter halos are better
represented by a flat core than by a cusp, \cite{Navarro03}, using
high resolution N-body simulations (down to 0.5 per cent of r200),
continue to defend that the fitting formula proposed by NFW provides a
reasonably good approximation to the density and circular velocity
profiles of individual halos.  Futhermore, they introduce a new
fitting formula with a steeper slope to alleviate some of the
differences with \cite{Moore99}. This seems at odds with observations,
especially of dwarf and LSB galaxies.
  
This is the third paper in a series on high resolution Fabry-Perot
spectroscopy of spiral galaxies. The first paper \citep[hereafter
paper I]{paperI} showed the necessity of optical integral field
spectroscopy to accurately determine the rotation curves in the inner
parts of spiral and dwarf galaxies. The main reason being that \hI
rotation curves are affected by beam smearing, a natural consequence
of their low spatial resolution
\citep{BegemanPhd,SwatersPhd,vdBosch00}. In combination with paper II
\citep{paperII}, paper I also brought to light the great sensitivity
of the mass distribution parameters to the inner rotation curve. The
ideal rotation curve is therefore a combination of high resolution
optical integral field spectroscopy and sensitive \hI radio
observations extending well beyond the luminous disk \citep[see
also][for high resolution CO data]{Sofue99}. Paper II has also shown
the difficulties in reconciling dwarf and late type galaxies rotation
curves with standard CDM N-body simulations
\citep[e.g.][]{NFW96,NFW97,Moore98,Moore99}. While those simulations do not
have the resolution to make definite predictions at small radii, an
extrapolation of the functional form derived at larger radii predicts
cuspy density profiles, implying much steeper inner slopes than
observed in these galaxies. Rotation curves of all galaxy types were
also shown to be more compatible with flat core density profiles.

\begin{table*}[hbt]
\caption{Parameters of the sample. Columns 4,5,6,10 and 11 are from the 
main reference unless otherwise specified.}
\label{tab:sample}  
\begin{tabular}{l l c c c c c c c c c c} \hline\hline
Name & Type & Dist. & D$_{25}$ & R$_{HO}$ & $\alpha^{-1}$ & i$^{p3}$ & PA$^{p3}$ & V$_{sys}$$^{p3}$ & M$_B$ & L$_B$ & Ref.\\
&        & Mpc & \arcmin & \arcmin & \arcsec & \degr & \degr & \kms & & $ 10^8$\lsol&\\ \hline
UGC 2259 & SBcd  & 9.6$^*$ & 2.6 & 1.9 & 28.2 & 41 & 165 & 581 & -17.03 & 9.7 & C88 \\
NGC 2403 & SABcd & 3.2$^{Fr}$ & 21.9 & 13.0 & 134 & 61 & 126 & 137 & -19.50 & 93.6 & C90  \\
NGC 6946 & SABcd & 6.2$^{Sh}$  & 11.5 & 7.8 & 115.2 & 38 & 60 & 47 & -21.38 & 530 & Be \\
NGC 5055 & SAbc  & 9.2$^*$  & 12.6 & 9.8 & 108.9 & 64 & 101 & 501 & -21.13 & 422 & Th \\
NGC 2841 & SAb   & 14.1$^{Ma}$ & 11.3 & 6.2 & 52 & 66 & 150 & 633 & -21.93 & 880 & Be \\ 
NGC 5985 & SABb  & 33.5$^*$ & 5.5 & 2.7 & 29.9 & 58 & 15 & 2515 & -21.74 & 738 & p3 \\ \hline 
\multicolumn{4}{l}{$^{p3}${\rule[0mm]{0mm}{4mm}This study}} & \multicolumn{8}{l}{Col 4: Diameter at the 25 B-mag/arcsec$^2$ isophote} \\
\multicolumn{4}{l}{$^{Be}${\cite{BegemanPhd}}} & \multicolumn{6}{l}{Col 5: Holmberg radius} \\ 
\multicolumn{4}{l}{$^{C88}${\cite{Car88}}}     & \multicolumn{6}{l}{Col 6: Scale length} \\
\multicolumn{4}{l}{$^{C90}${\cite{Car90}}}     & \multicolumn{6}{l}{Col 7: Inclination}\\
\multicolumn{4}{l}{$^{Fr}${\cite{Freedman90}}} & \multicolumn{6}{l}{Col 8: Position angle of the major axis} \\
 \multicolumn{4}{l}{$^{Ma}${\cite{Macri01}}} & \multicolumn{6}{l}{Col 9: Systemic heliocentric velocity} \\
\multicolumn{4}{l}{$^{Sh}${\cite{Sharina97}}} & \multicolumn{6}{l}{Col 10: Absolute B magnitude} \\
\multicolumn{4}{l}{$^{Th}${\cite{Thornley97}}} & \multicolumn{6}{l}{Col 11: Total B luminosity} \\          
\multicolumn{4}{l}{$^{*}${H$_0$ = 75\kms/Mpc}}
\end{tabular}  
\end{table*}

The first step in showing the reality of this discrepancy is to
eliminate the known possibilities of systematic observational biases.
Two classes of error can contribute to the underestimation of the
velocities, hence of the computed halo density, in the inner parts of
spiral galaxies. The prime culprit in radio observations is the ``beam
smearing'' effect due to the relatively low angular resolution of 21
cm data. Typically the radio beam is too large to resolve the inner
disk velocity gradient. Combining the \hI density gradient with the
true velocity gradient inside a beam width will most likely lead to
underestimate the velocity at a given radius.

\ha observations always easily reach an angular resolution where any
spatial resolution effect can be neglected. A less often commented
source of uncertainties though is found in long slit observations,
where most of the \ha data come from.  \cite{Schommer93} was the first
large scale publication of rotation curves derived from Fabry-Perot
data.  They clearly state the inherent advantages of 2D spectroscopy
over long slit observations.  Beauvais and Bothun (1999, 2001) were
among the first to do Fabry-Perot imaging spectroscopy of nearby,
large angular size late type spirals for the express purpose of
performing precision mass modeling.  They detailed the advantages that
2D velocity fields have over 1D slits.  Beauvais and Bothun (2001)
discussed precision quality rotation curves and optimal rotation curve
fitting One can point to the lack of 2D coverage that makes the
alignment of the slit crucial to retrieve the real kinematics of a
galaxy.  First, a photometrically determined inclination can be a
major source of uncertainty.  Second, missing the kinematical center,
which is not always the photometric center (paper II) can lead to an
underestimation of the rotational velocities. Also, just a degree or
two between the slit and the galaxy position angle for highly inclined
galaxies can cause the highest velocity regions along the line of
nodes to be missed. This in turn leads to underestimate the rotational
velocities.  In a series of papers started in 1992 \citep[referenced
in][]{Amram96}, a large sample of rotation curves of nearby galaxies
in clusters, derived from Fabry-Perot, have also shown the
observational biases avoided when using 3D data cubes instead of long
slits data. In addition, the presence of a bar would hardly be noticed
in long slit data and its effect would most probably be confounded
with the rotational kinematics, while 2D velocity fields allow to
disentangle circular from radial motions.

This paper extends our sample of multi-wavelength velocity fields to
earlier morphological types by adding six spiral galaxies to the
optical rotation curves sample with available 21 cm data.  Contrary to
dark matter dominated dwarf spirals and to luminous matter dominated
early type spirals, the stellar disk and dark halo of intermediate
type have comparable contributions to the mass inside the Holmberg
radius. Therefore, a slight change of the inner slope of the rotation
curve can significantly reduce or increase the disk contribution,
whether it is maximal or best--fitted. This change can induce a
dramatic difference in the dark matter distribution by increasing or
limiting its inner density, as seen in the case of NGC 5585 in
\cite{paperI}. The halo core radius and central density are indeed
very sensitive to the exact baryonic contribution at low radii.  This
is true for halos of all shapes, but more critical for cuspy halos,
which often have problem to accommodate for the presence of even a
small disk in later type spirals \citep{paperII}.

Section \ref{sec:sample} briefly describes the sample while section
\ref{sec:fpdata} explains the Fabry--Perot observations and data
reduction, and presents the main features of each galaxy. Computed \ha
velocity fields and rotation curves are also described therein, and
compared to \hI rotation curves. Dynamical analysis and mass models
will be given in a forthcoming paper (paper IV). Section
\ref{sec:discussion} presents a discussion of the main results and the
main conclusions.

\section{The sample}
\label{sec:sample}

Table \ref{tab:sample} summarizes the optical parameters of each
galaxy in the sample. Neutral hydrogen kinematics and mass models of
UGC 2259 and NGC 6946 have been studied by \cite{Car88,Car90} while
NGC 2403 and NGC 2841 are part of Begeman's thesis
(\citeyear{BegemanPhd}). All these studies present rotation curves
based on 21 cm observations from the Westerbork Synthesis Radio
Telescope (WSRT). NGC~5985 is part of the Westerbork \hI survey of
spiral and irregular galaxies\footnote{www.astro.rug.nl/$\sim$whisp}
\citep[WHISP, see][]{Swaters00}. An \hI rotation curve derived from
these data is presented in this paper, but the details on the \hI
observations for this galaxy will be published elsewhere.  An \hI
rotation curve of NGC 5055, based on VLA observations, has been
published by \cite{Thornley97}.

This sample expands the range of morphological types from the very
late types of paper II to earlier (Sc to Sb) types. The GHASP
survey\footnote{http://www-obs.cnrs-mrs.fr/interferometrie/ghasp.html}
\cite{Garrido02}, an ongoing Fabry-Perot survey will expand this
sample to about 200 spiral galaxies.

\section{Fabry--Perot observations: data acquisition and reduction}  
\label{sec:fpdata}  

\begin{table}[htb]
\caption{Parameters of the Fabry--Perot observations.}
\begin{minipage}{\hsize}
\renewcommand{\footnoterule}{\rule{0pt}{0pt} \vspace{-2mm}}
\renewcommand{\thefootnote}{\alph{footnote}}
\begin{center}
\label{tab:fpobs4}  
\begin{tabular}{lr} \hline\hline
Date of observations        & September 97, March 98            \\  
Telescope                   &            3.6\,m CFHT\footnotemark[1]        \\  
Instrumentation:                                                        \\  
~~~Instrument &               MOSFP                           \\  
~~~CCD detector    &       2048\,$\times$\,2048, $\sigma$ = 8\,e$^{-1}$         \\  
~~~Filters:                                                           \\
~~~~~~~~  UGC 2259:    &  $\lambda_0$ = 6575\,\AA,~\, $\Delta \lambda$ = 13\,\AA \\  
~~~~~~~~  NGC 2403:    &  $\lambda_0$ = 6568\,\AA,~\, $\Delta \lambda$ = 12\,\AA \\  
~~~~~~~~~  NGC 6946:    &  $\lambda_0$ = 6566\,\AA,~\, $\Delta \lambda$ = 12\,\AA \\  
~~~~~~~~~  NGC 5055 blue:    & $\lambda_0$ = 6575\,\AA, $\Delta \lambda$ = 13\,\AA \\
~~~~~~~~~~~~~~~~~~~~~~~~~ red: &  $\lambda_0$ = 6586\,\AA,~\, $\Delta \lambda$ = 12\,\AA \\
~~~~~~~~~  NGC 2841 blue:    &  $\lambda_0$ = 6577\,\AA,~\, $\Delta \lambda$ = 10\,\AA \\  
~~~~~~~~~~~~~~~~~~~~~~~~~ red   &  $\lambda_0$ = 6587\,\AA,~\, $\Delta \lambda$ = 11\,\AA \\  
~~~~~~~~~  NGC 5985:    &  $\lambda_0$ = 6621\,\AA,~\, $\Delta \lambda$ = 26\,\AA \\  
~~~~Fabry--Perot unit   &  Scanning QW1162 (CFHT1)     \\  
~~~~Calibration lamp       &        Neon ($\lambda$ = 6598.95\,\AA)    \\  
~~~~Interference order     &            1155 @   
                        $\lambda_{\scriptscriptstyle N\!E\!O\!N}$       \\  
~~~Mean Finesse           &    12                             \\  
Duration:                                                               \\
~~~UGC 2259:    &  8\,min/channel (3.6\,h) \\
~~~NGC 2403:    &  7.34\,min/channel (2.9\,h) \\  
~~~NGC 6946:    &  7.5\,min/channel (3.4\,h) \\  
~~~NGC 5055 blue:    &  9.42\,min/channel (3.8\,h) \\ 
~~~~~~~~~~~~~~~~~ red:       &  6\,min /channel (2.7\,h) \\ 
~~~  NGC 2841\footnotemark[1]:    &  5\,min/channel (2\,h) \\  
~~~  NGC 5985:    &  7.7\,min/channel (3.1\,h) \\  
Spatial Parameters          &                                           \\  
~~~Field size             &            8.5$'$\,$\times$\,8.5$'$       \\  
~~~Pixel scale            &            0.44\arcsec\,pix$^{-1}$        \\
Spectral parameters         &                                           \\
~~~Free spectral range at \ha    &     5.66\,\AA\ (258\,km\,s$^{-1}$)         \\  
NGC 2403, 5055, 5985: &                                         \\
~~~Number of channels     &                             24            \\  
~~~Sampling               &     0.24\,\AA\ (10.8\,km\,s$^{-1}$)/channel \\
UGC 2259 and NGC 6946:      &                                           \\   
~~~Number of channels     &                             27            \\  
~~~Sampling               &     0.21\,\AA\ (9.6\,km\,s$^{-1}$)/channel \\ 
NGC 2841\footnotemark[1]:           &                                   \\
~~~Free spectral range at \ha   &     8.30\,\AA\ (378\,km\,s$^{-1}$)  \\  
~~~Number of channels     &                             24            \\  
~~~Sampling               &     0.35\,\AA\ (15.8\,km\,s$^{-1}$)/channel \\ \hline
\end{tabular}  
\end{center}
\footnotetext[1]{NGC 2841 was reobserved at the Observatoire de Haute Provence. See text.}
\end{minipage}
\end{table}

All the Fabry-Perot observations have been initially made at the
Canada--France--Hawaii Telescope (CFHT) in September 97 and March 98
using the CFHT1 high resolution Fabry-Perot etalon installed in the
Multi--Object Spectrograph (MOS) focal reducer. A narrow--band filter
($\Delta \lambda \simeq$ 12\,\AA), centered around the systemic
velocity of the observed object, was placed in front of the etalon.
The available field with negligible vignetting was $\approx$
8.5\arcmin $\times$ 8.5\arcmin~, with 0.44\arcsec\, pix$^{-1}$. The
free spectral range of 5.66\,\AA\, (258 \kms) was scanned in 27 (+1
overlapping) channels for UGC 2259, NGC 6946 and the blue side of NGC
5055 (see section \ref{sec:5055}), giving a sampling of 0.21\,\AA\,
(9.6 \kms) per channel. For the rest of the sample, the objects were
scanned in 24 channels, for a sampling of 0.24\,\AA\, (10.8 \kms).
Due to the use of an aging filter which has apparently blueshifted
from its quoted central wavelength, the fluxes from the receding half of
NGC 2841 and NGC 5055 were almost totally blocked, and these
galaxies had to be re-observed. The details are given in the relevant
sections.

Following normal de--biasing and flat--fielding with standard IRAF
procedures, a robust 3-D cosmic--ray removal routine, that tracks
cosmic rays by spatial (pixel--to--pixel) and spectral
(frame--to--frame) analysis, was applied to every data cube. When
necessary, ghost reflections were then removed using the technique
described in Paper I. A neon line ($\lambda$6598.95 \AA) was used for
absolute wavelength calibration at each pixel.

\begin{figure*}
        \resizebox{\hsize}{!}{\includegraphics{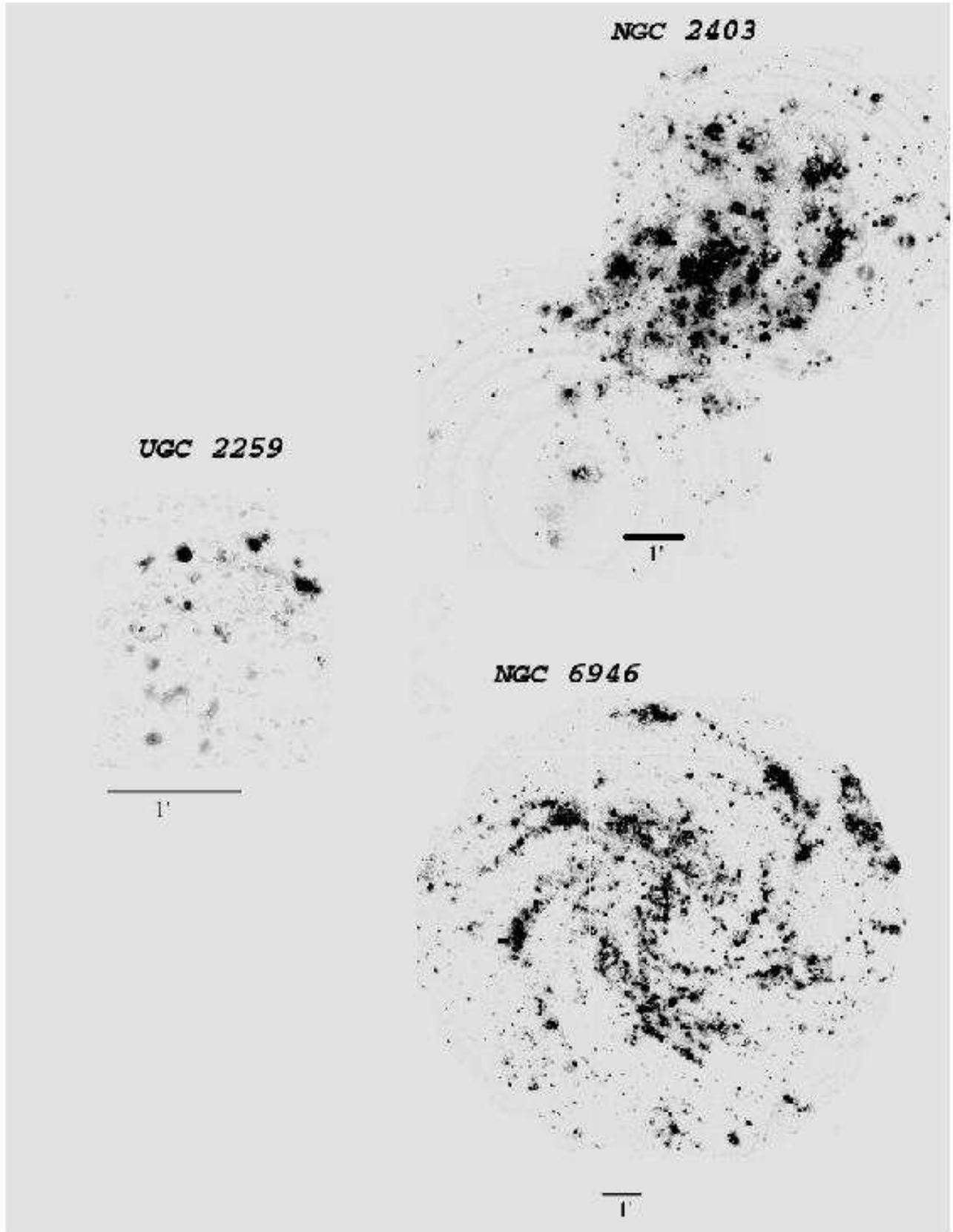}}
        \caption{Monochromatic images of the \ha flux of UGC 2259, NGC 2403 and NGC 6946. North is up, East is left.}
        \label{fig:mono1}
\end{figure*}

\begin{figure*}
        \resizebox{\hsize}{!}{\includegraphics{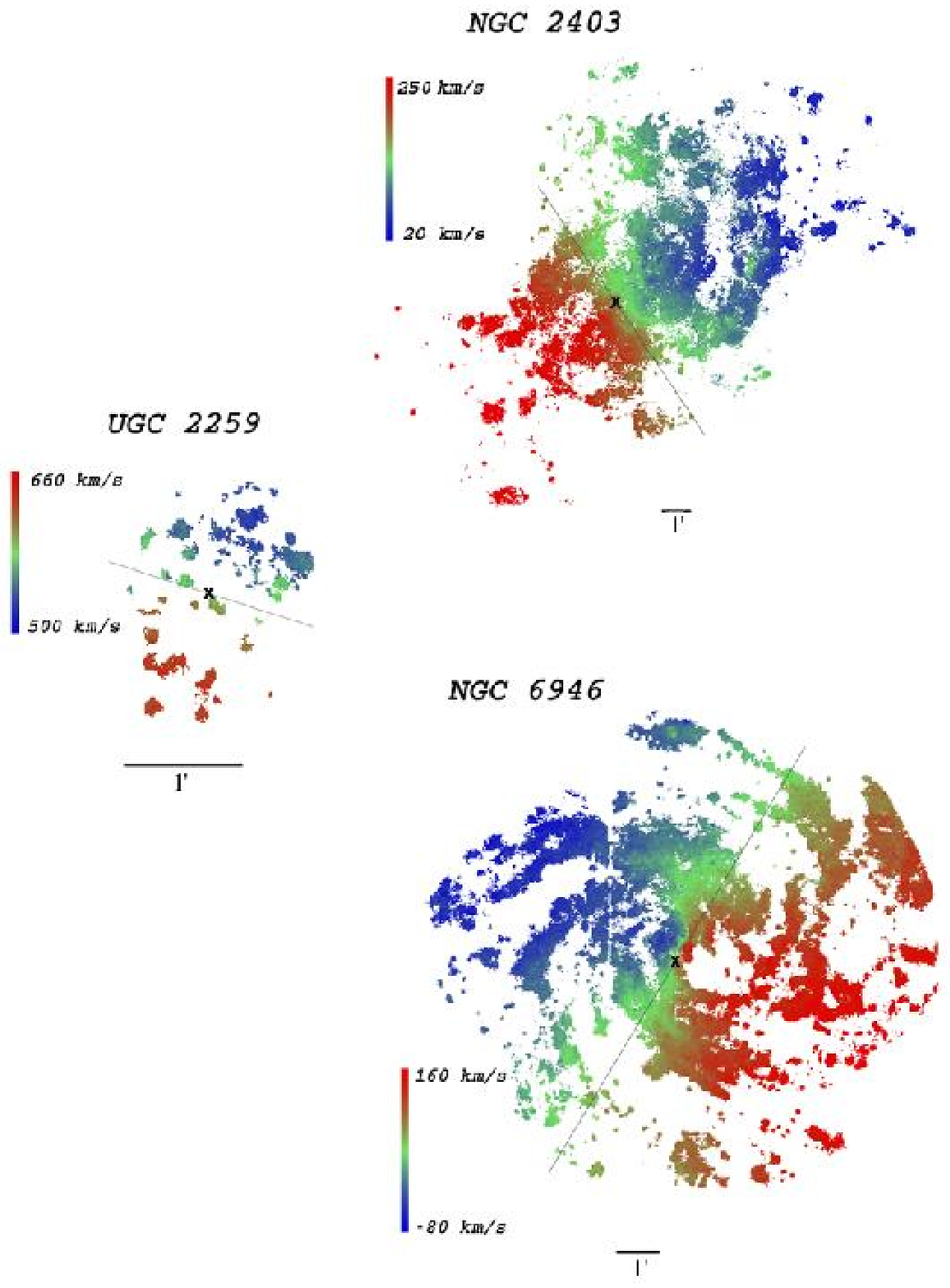}}
        \caption{Velocity fields of the same galaxies.The X and the
        grey line indicate the kinematic center and the axis of
        separation between the approaching and receding sides.}
        \label{fig:vf_1}
\end{figure*}

\begin{figure*}
        \resizebox{\hsize}{!}{\includegraphics{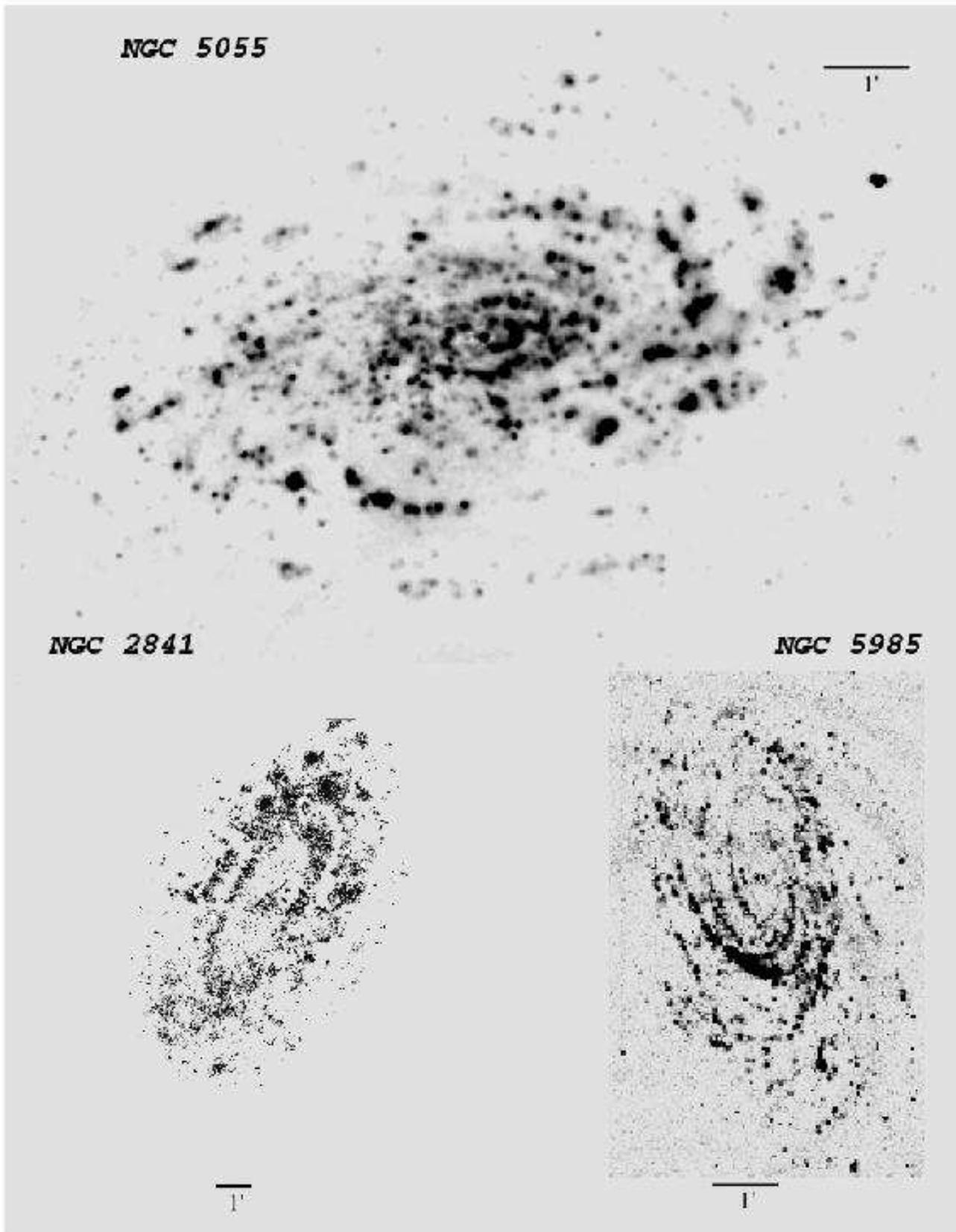}}
        \caption{Monochromatic images of the \ha flux of NGC 5055, NGC 2841 and NGC 5985}
        \label{fig:mono2}
\end{figure*}

\begin{figure*}
        \resizebox{\hsize}{!}{\includegraphics{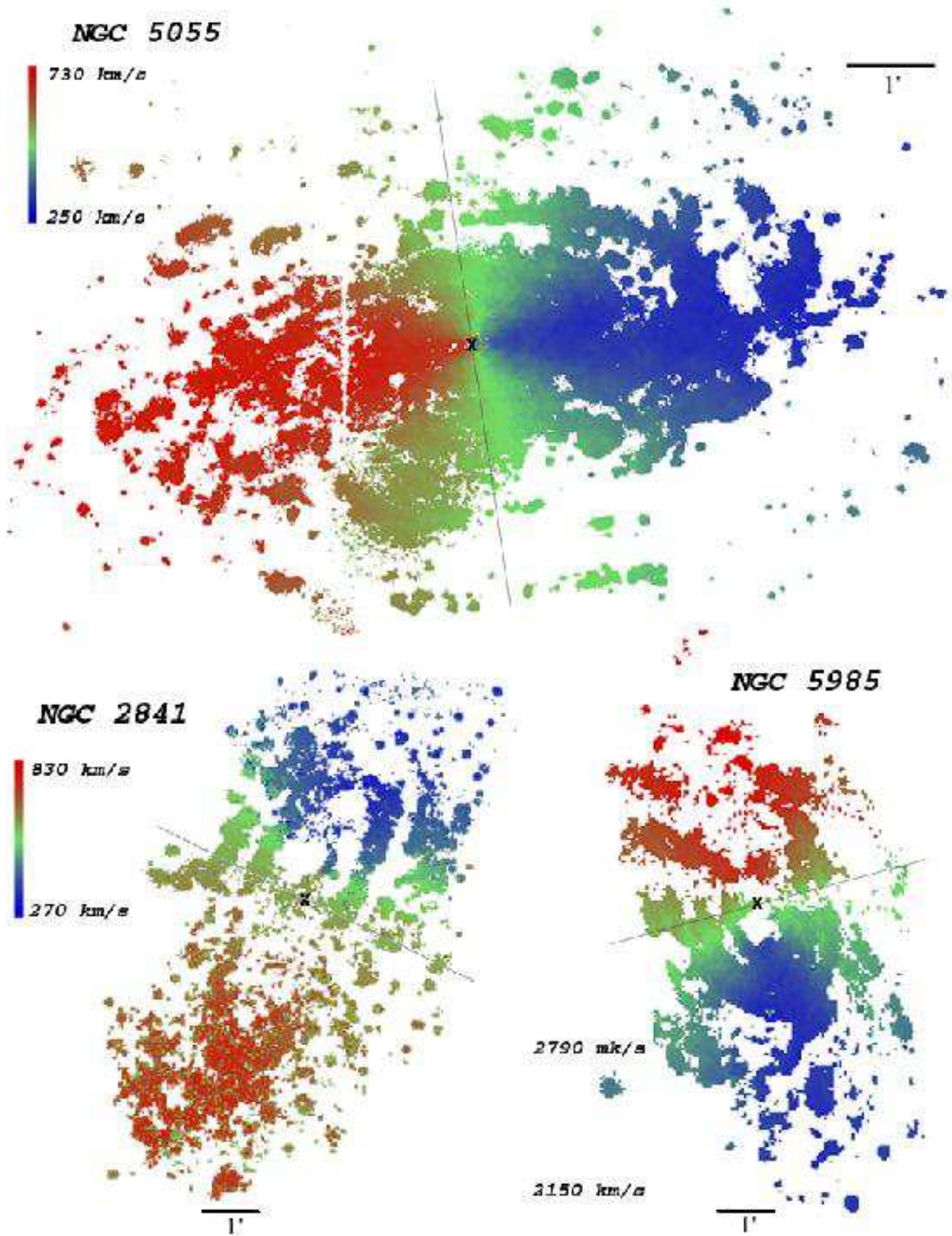}}
        \caption{Velocity fields of the same galaxies.}
        \label{fig:vf_2}
\end{figure*}

The signal measured along the scanning sequence was split into two
parts: (1) an almost constant level produced by the continuum light in
a narrow passband around H$\alpha $ (hereafter referred to as
continuum map) and (2) a varying part produced by the H$\alpha $ line
(hereafter referred to as H$\alpha$ emission line map \textit{or}
monochromatic map). The continuum level was taken to be the mean of
the channel which do not contain the line.  The H$\alpha$ integrated
flux map was obtained by integrating the monochromatic profile in each
pixel above the threshold defined by the continuum level.  The
scanning of the interferometer sufficiently samples the PSF of the
instrument (the Airy function convolved by the surface and
transmission defects) and covers the free spectral range through 24 to
27 scanning steps.  When the profiles are structureless and the S/N
high, the lines could be successfully fitted by the convolution of the
observed PSF (given by the narrow Neon-6598.95nm emission line) and a
single Gaussian function.  When the profile is more complex, multiple
Gaussian components are needed.  Nevertheless, since the number of
scanning steps and the baseline of the continuum emission are
relatively low, and the structure of the profile is often complex
(asymmetries, multiple components, low S/N), we do not fit a function
to the profile to extract the first order momentum.  Instead, the
heliocentric radial velocity for a given pixel is directly given by
the position of the barycentre of the line. Furthermore, we do not
have to make assumptions on the fit used when the data are dominated
by Poisson' or receptor' noise at low S/N levels.  However, at high
S/N, since we have a good knowledge of the PSF, the barycentre of the
emission line profile may be measured with an accuracy much better
than the sampling step, hence giving a precision of about 3 km/s for a
S/N=5.  For each pixel, the S/N level is given by the y-axis of the
barycentre of the line normalized by the r.m.s of the continuum.  When
the S/N of each individual pixel was not high enough to derive a
radial velocity, two different Gaussian smoothings were performed on
the cubes ($\sigma=2.6\arcsec$ and 4\arcsec) in order to get
sufficient signal--to--noise throughout the images. Three velocity
maps (one for each smoothed cube and for the original) were then
obtained from the intensity weighted means of the H$\alpha$ peaks at
each pixel.  A final variable resolution velocity map was constructed
keeping higher resolution where the signal-to-noise makes it possible.
Figs.  \ref{fig:mono1} and \ref{fig:mono2} show the monochromatic
images of the galaxies while Figs. \ref{fig:vf_1} and \ref{fig:vf_2}
display their velocity fields\footnote{All the velocity maps are
  available at www.astro.caltech.edu/$^{\sim}$sbo/vf.html}.
Table~\ref{tab:fpobs4} lists the main observing parameters.

Rotation curves were derived using two different methods: velocity
dispersion minimization \citep[tilted-ring model:][]{BegemanPhd,
  Cote91} and rotation curve symmetry optimization by comparing the
approaching and receding sides. The first method is more
precise and allow to model a warp disk, while the second is
more robust and is useful for galaxies with low S/N and patchy
velocity coverage. In all cases all the parameters (systemic velocity,
kinematical center, inclination, and position angle) were initially
free to vary.

To date, no convention on the way to represent the errors on rotation
curves exists in the literature. Error bars are often simply given as
the velocity dispersion in the ring used at each radius. However, the
warm gas tends to be found in more disturbed environments than the
cold gas. Turbulence, local density variations (like spiral arms) and
winds from stars and supernovae of the young stellar forming regions
in which the gas is ionized, increase its dispersion. This can lead to
the paradox where fewer points (as in long--slit observations)lead to
lower dispersion and to smaller error bars. As a more direct probe of
the uncertainties on the measured potential, the difference between the
two sides of the galaxy is instead sometimes used. Some authors add
the error due to uncertainties on inclination and/or position angles.

\begin{figure*}[htb]
\includegraphics[angle=0,width=\textwidth]{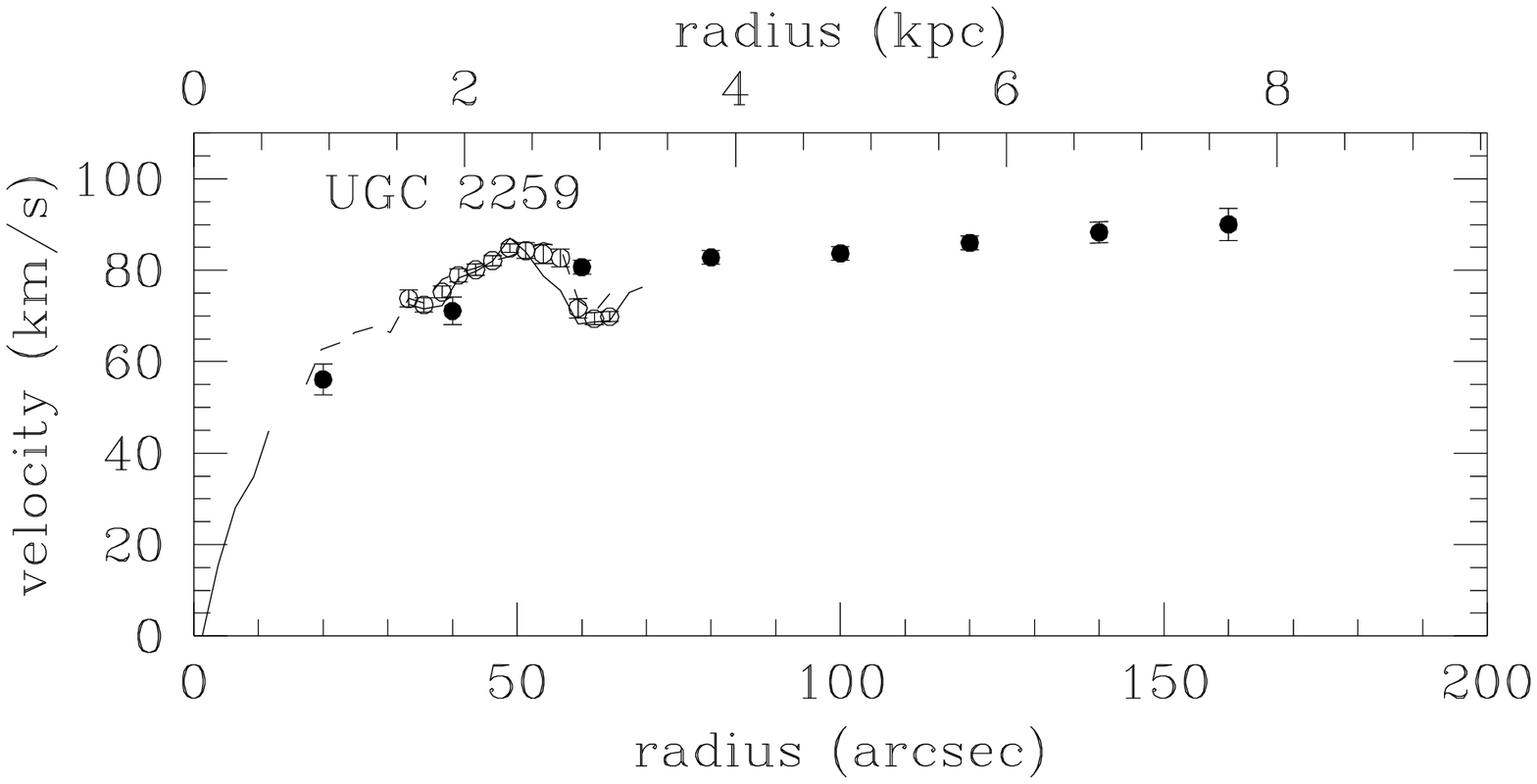}
\caption{\ha rotation curve of UGC 2259 (open circles). Approaching and
receding sides from \ha data are indicated respectively by the dashed
and continuous lines. The filled circles represent the \hI rotation
curve from \cite{Car88}.
\label{fig:rc_2259}}
\end{figure*}

Since the purpose of this study is to present the data in an unbiased
way, the sources of errors are given separately. The error bar itself
gives the error on the mean in each ring ($\sigma/\sqrt{N}$) while the
solutions for each side of the galaxy are represented by continuous
lines for the receding side and dashed lines for the approaching.
Errors due to inclination uncertainties are given by $V (\sin
i+\epsilon - sin i-\epsilon)$ where $\epsilon$ is around 2 for all the
galaxies. Other parameters (Table \ref{tab:sample}) have low
uncertainties and contribute marginally to the velocity error budget.

\begin{figure*}[htb]
\includegraphics[angle=0,width=\textwidth]{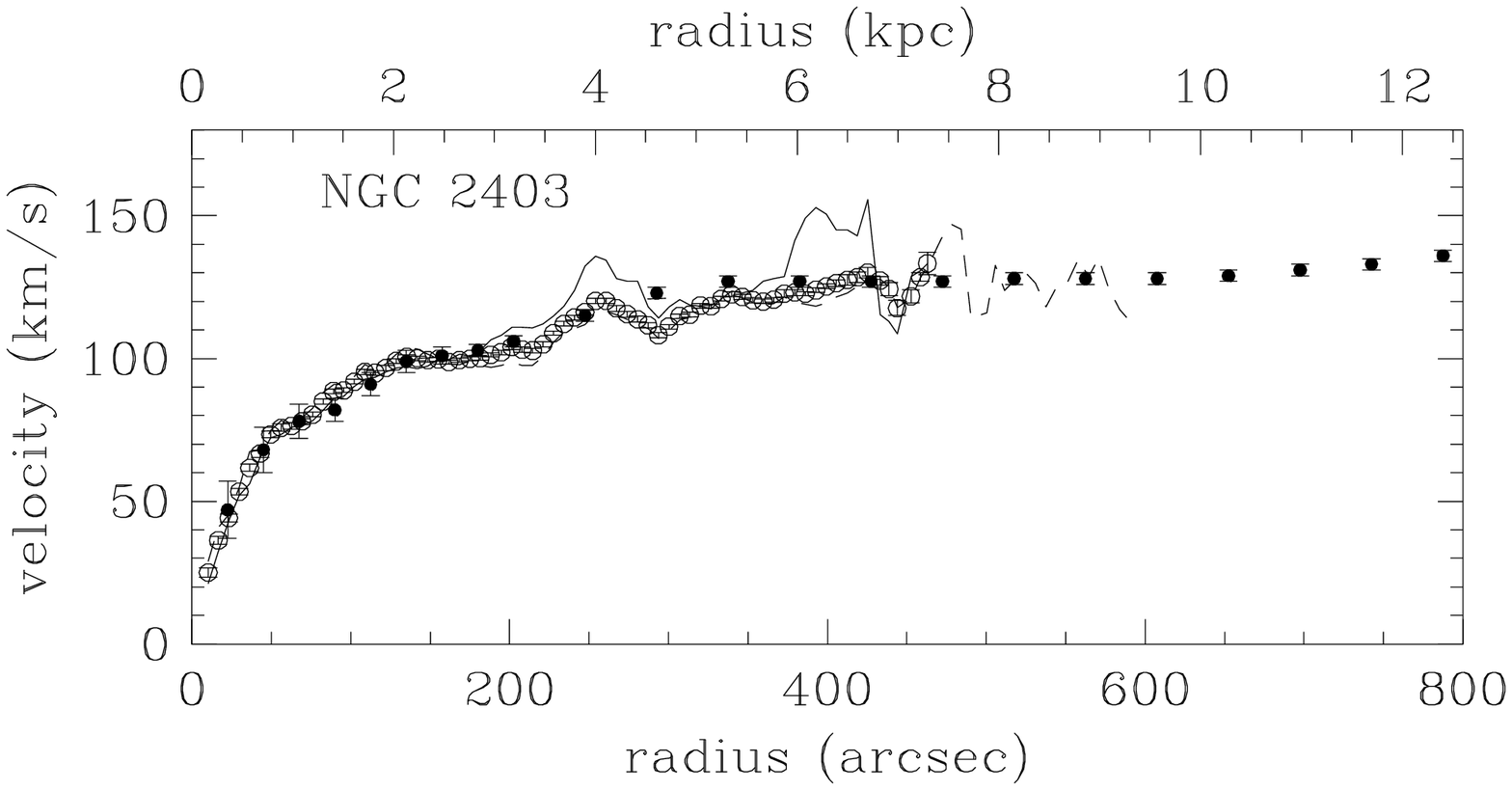}
\caption{\ha rotation curve of NGC 2403 (open circles). Approaching and
receding sides from \ha data are indicated respectively by the dashed
and continuous lines. The filled circles represent the \hI rotation
curve from \cite{BegemanPhd}.
\label{fig:rc_2403}}
\end{figure*}

\subsection{UGC 2259}
   
This small galaxy was observed in September 97 by a fairly photometric
night. Eight minutes integration were spent at each channel position
and no photometric corrections were needed.

Like in the case of IC 2574 (paper II), dispersion minimizing methods
as in ROCUR are hardly applicable to determine the rotation curve
because of the incomplete coverage of the velocity field. The rotation
curve was thus found by comparison of the two sides of the galaxy and
analysis of the residual velocity field using the ADHOC
package\footnote{http://www.oamp.fr/adhoc/adhoc.html}. The
inclination, position angle and systemic velocity were fixed at
41\degr, 158\degr~ and 581\kms, in fairly good agreement with
\cite{Car88}. The \ha coverage of the field is patchy and mainly
concentrated in the center and in the two spiral arms. The net effect
is to have only the receding side covering the innermost part of the
rotation curve. However, because the two sides cover the outer parts
and agree fairly well, this lack of coverage should only affect the
determination of the kinematical center, adding some uncertainty to
the rotation curve. The resulting optical rotation curve at
2.6\arcsec~resolution is given in Table~\ref{tab:rc_2259_ha} while
Fig. \ref{fig:rc_2259} presents it in combination with the \hI
rotation curve.

As one can see in Fig.\ref{fig:vf_1}, only a few clumps of ionized gas
reveal the very inner kinematics of the galaxy. The solid body rise
can now be observed within a few hundreds of parsecs on the receding
side. The \ha curve is higher than the \hI by 5 to 10
\kms up to a radius of 50\arcsec. Even if this is an indication of the
presence of beam smearing in the uncorrected \hI observations, we
cannot exclude non-circular motions due to the lack of coverage.

The most prominent feature of the \ha rotation curve is the big dip
around 60\arcsec. This is not observed in the \hI curve. It can be
seen in the velocity field that the location of this feature
corresponds to the two great spiral arms, clearly identified in the
optical image \citep[see e.g.][]{Car88}. Spiral arms are known to create
such effect, especially on the ionized gas \citep[e.g.][]{Thornley97}.

Fabry-Perot observations of UGC 2259 by \cite{Gonzalez91} using the
brightest \hII regions and the new data are consistent within the
error bars.

In conclusion, even if the kinematical parameters derived for UGC 2259
reach a reasonable level of confidence, the asymmetric distribution of
H$\alpha$ emission and the relatively low covering factor weaken this
galaxy contribution as a test case.

\subsection{NGC 2403}

The object being larger than the available field, two data cubes were
acquired during the run of March 98. The same parameters were used for
both fields and 440 seconds were spent on each of the 24 channels of
each data cube. After the standard data reduction, the final maps
(velocity, continuum and monochromatic) were joined using stars in the
overlapping region. The coverage of the field was this time complete
enough to use ROCUR with bins of 20 arcsec. However, since no
significant variations of inclination or position angle were found,
ADHOC was used to get a higher sampling. The inclination was found to
be 61\degr, the position angle 126\degr and the systemic velocity 137
\kms, at 3 \kms of Begeman's value. The shape of the rotation curve is
independent of the resolution used and bins of 5 pixels (6.6 \arcsec)
offer the clearest picture.  Fig. \ref{fig:rc_2403} shows the final
rotation curve. The \hI curve is also plotted for comparison.  The
Fabry--Perot data show a remarkable agreement between the approaching
and receding sides up to 200\arcsec. At a greater radius, the number
of points on the receding side drops sharply (see Table
\ref{tab:rc_2403_ha}) and the curve then relies mainly on the
approaching side.

\begin{figure*}[htb]
\includegraphics[angle=0,width=\textwidth]{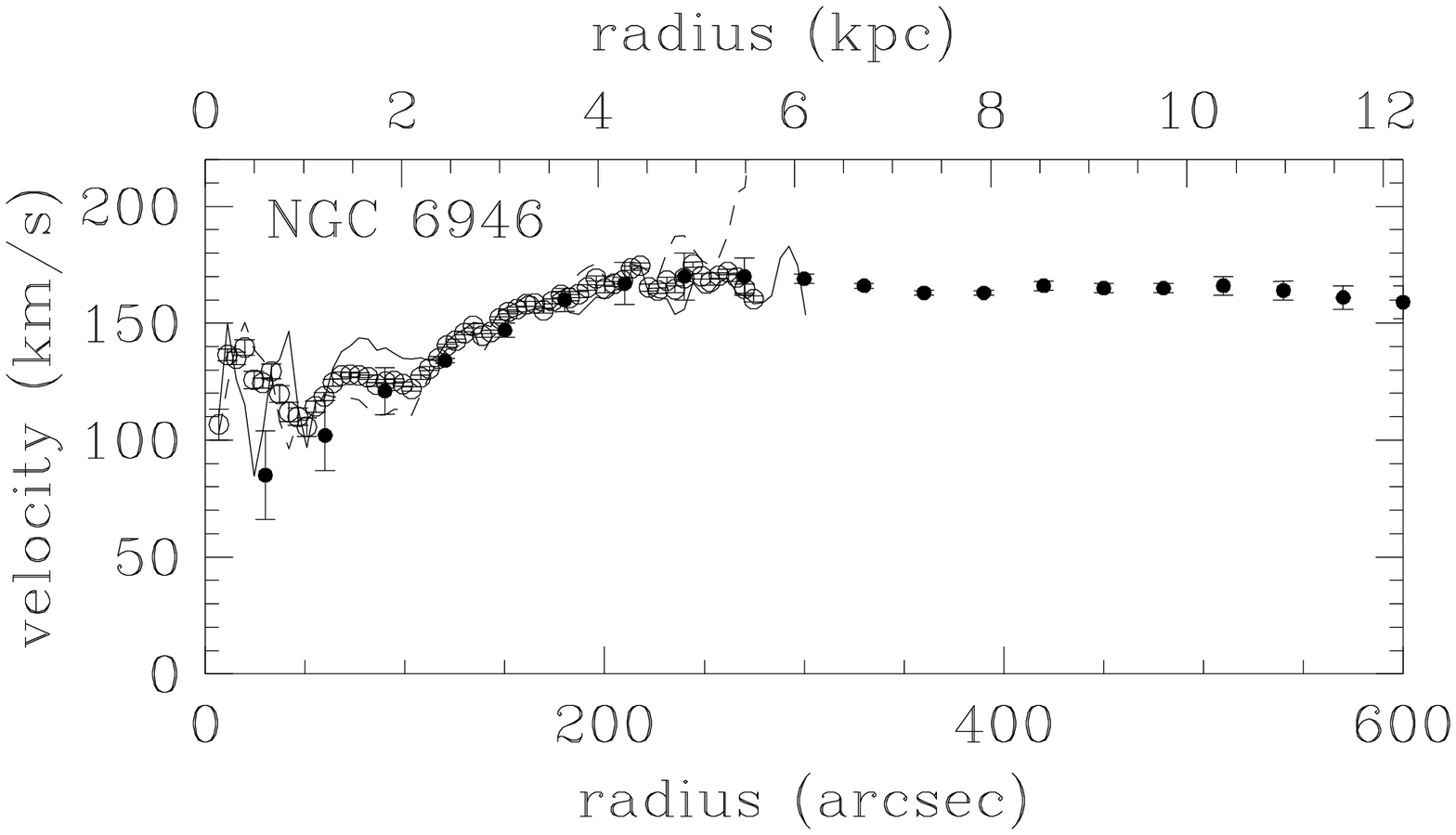}
\caption{\ha rotation curve of NGC 6946 (open circles). Approaching and
receding sides from \ha data are indicated respectively by the dashed
and continuous lines. The filled circles represent the \hI rotation
curve from \cite{Car90}
\label{fig:rc_6946}}
\end{figure*}

The inner limit of the big external arm is located around 200\arcsec\,
and the effect of the arm can clearly be seen on both sides of the
rotation curve. The full resolution velocity map gives an even clearer
picture as it shows the effect of the arm at different galactic
longitudes. The south--eastern arm is less obvious on the
monochromatic image and seems dominated by a few giant \hII
regions. This explains the bumpy appearance of the receding curve
above 200\arcsec~where each \hII region produces a bump and increase
the dispersion. Only the approaching side is seen above 470\arcsec~and
relies only on a few
\hII regions.

\subsection{NGC 6946}

The central part of this bright Scd galaxy suffers from a lack of \ha
emission. Despite the 3.5 hours of observation that otherwise gave the
sensitivity to reach down to the smaller \hII region and the diffuse
ionized gas, few \ha photons were collected up to 50\arcsec~ in radius. The
velocity dispersion is high in that region and makes the true
kinematics hard to retrieve. This appears clearly in the rotation
curve presented in Fig. \ref{fig:rc_6946} and in Table
\ref{tab:rc_6946_ha}. 

Between 70\arcsec~and 100\arcsec, the rotation curves on both sides
are smooth but present a big asymmetry, indicative of non circular
motions, most probably gas movement along a central bar
\citep{Bonnarel88}. The presence of dust \citep{Roussel01} explains
the lack of \ha emission and is compatible with a mild starburst
occurring at the center. This starburst is likely to play a role in the
high velocities observed near the center, considering that the
relatively low mass of the molecular gas, a few percent of the stellar
mass \citep{Israel01}, could not explain such high rotational
velocities in the inner part.

The rest of the rotation curve (resolution of 4.4\arcsec) matches the
\hI curve. The systemic velocity is found to be 47 \kms in agreement
with \cite{Car90} but not consistent with the 38\kms found by
\cite{Bonnarel88}. However, the inclination and position angles are
less constrained due to the uncertainties caused by the global
asymmetry. They are compatible with both Bonnarel (i = 32\degr and PA
= 58\degr) and \citeauthor{Car90} (i = 38\degr and PA = 60\degr). They
were fixed to the latter for ease of comparison.  It has to be noted
that the uncertainty on the inclination can affect the velocities
dramatically (as $\sin i$). For example, \cite{Sofue96} uses an
optically determined inclination of 30\degr\ in conjunction with his
CO data along the major axis and ends up with velocities 20\%
higher. That recalls the importance of having a complete 2-D
coverage to allow an independent determination of optical parameters.

\subsection{NGC 5055}
\label{sec:5055}
                                                                           
\begin{figure*}[htb]
\includegraphics[angle=0,width=\textwidth]{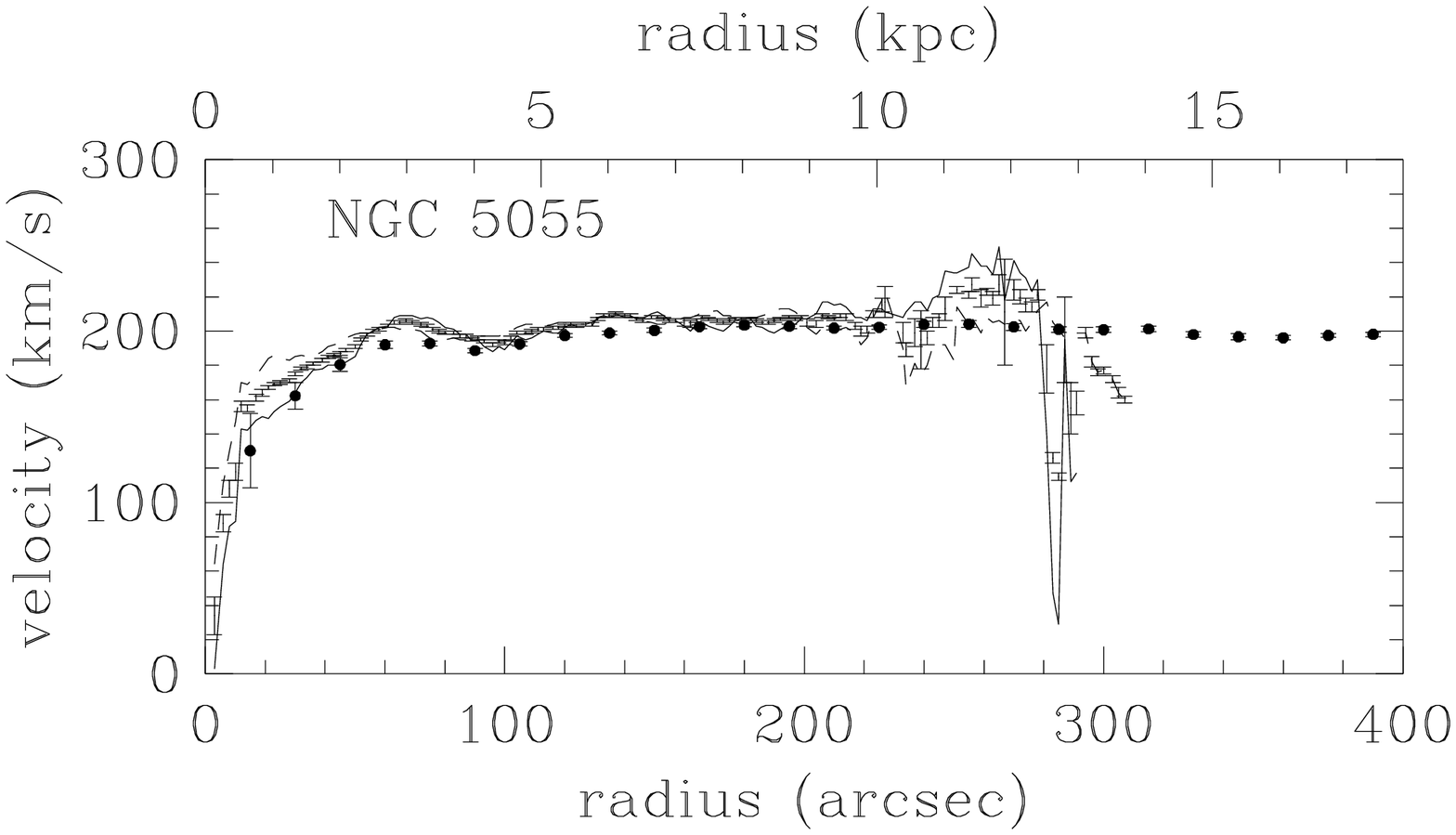}
\caption{\ha rotation curve of NGC 5055 (error bars alone). Approaching and
receding sides from \ha data are indicated respectively by the dashed
and continuous lines. The filled circles represent the \hI rotation
curve from \cite{Thornley97}.
\label{fig:rc_5055}}
\end{figure*}

This object was first observed in March 98 but the interference filter
used had blueshifted as explained in section \ref{sec:fpdata}. To
complete the data set, it has been observed again in April 99 using a
redder filter. Both sets of data were analyzed following the standard
procedure and two full resolution data cube were produced. They have
then been combined using field stars and their spectral zero point
matched with high accuracy in overlapping regions.

For the majority of bright galaxies where the kinematics have been
investigated in detail, the optical and dynamical centers do coincide
\citep{BegemanPhd}.  For some late-type barred galaxies the shifts
between both center determinations could be larger than 1 kpc (e.g.
\citep[e.g.][]{deVaucouleurs72, Beauvais01, Weldrake03}. For low
surface brightness galaxies, due do their faintness, the determination
of the morphological center could in itself be a problem, nevertheless
\cite{deBlok03} conclude that the offsets between optical and
dynamical centers are small. We have tested for NGC 5055 the influence
of a ``kinematic sloshing of the center of mass'' on the spatial scale
of 500 pc and we conclude that whether inclination and position angles
were kept fixed at the value derived by \cite{Thornley97} or let free
to vary (in which case the averages are 64\degr~and 99\degr), the \ha
rotation curve (Fig.  \ref{fig:rc_5055} and \ref{tab:rc_5055_ha})
ranges systematically from 2 to 11 \kms above the \hI curve. This fact
is hard to understand especially in view of the otherwise general good
agreement between the two curves.  \cite{Pismis95} published a
long--slit based rotation curve that tends to be slightly lower than
the present Fabry-Perot curve. However, because of its limit in radius
(70\arcsec), it is hardly reliable to solve the discrepancy.  In an
early paper, \cite{vanDerKruit78} derived an optical rotation curve
from the inner 60 arcsec regions of NGC 5055.  Taking into account the
difference of inclination the agreement with our rotation curve is
reasonable.

Two important differences exist between the \hI and \ha curves. First,
the central \ha points are up to 30 \kms higher than their \hI
counterparts.  Second, it can be noticed that the amplitude of the
velocity variations due to the passage of the spiral arms (at radii
around 60 \arcsec\ and 120\arcsec) are much more pronounced in the \ha
data than in \hI.  The velocity difference can be in part naturally
associated with beam smearing in the 21 cm observations. However, its
amplitude, the strength of the response to arm, and especially the
difference between approaching and receding sides show an important
component of streaming motion, in agreement with the finding of
\cite{Thornley97}. Averaging velocities from both sides helps removing
the effect of this non-circular motion and retrieving the equivalent
rotational kinematics. Nonetheless, this galaxy is a good illustration
of the ionized gas sensitivity to its environment, due to the short
time scale of kinematic feedback from star formation processes.

\begin{figure*}
  \centering \includegraphics[angle=0,width=15cm]{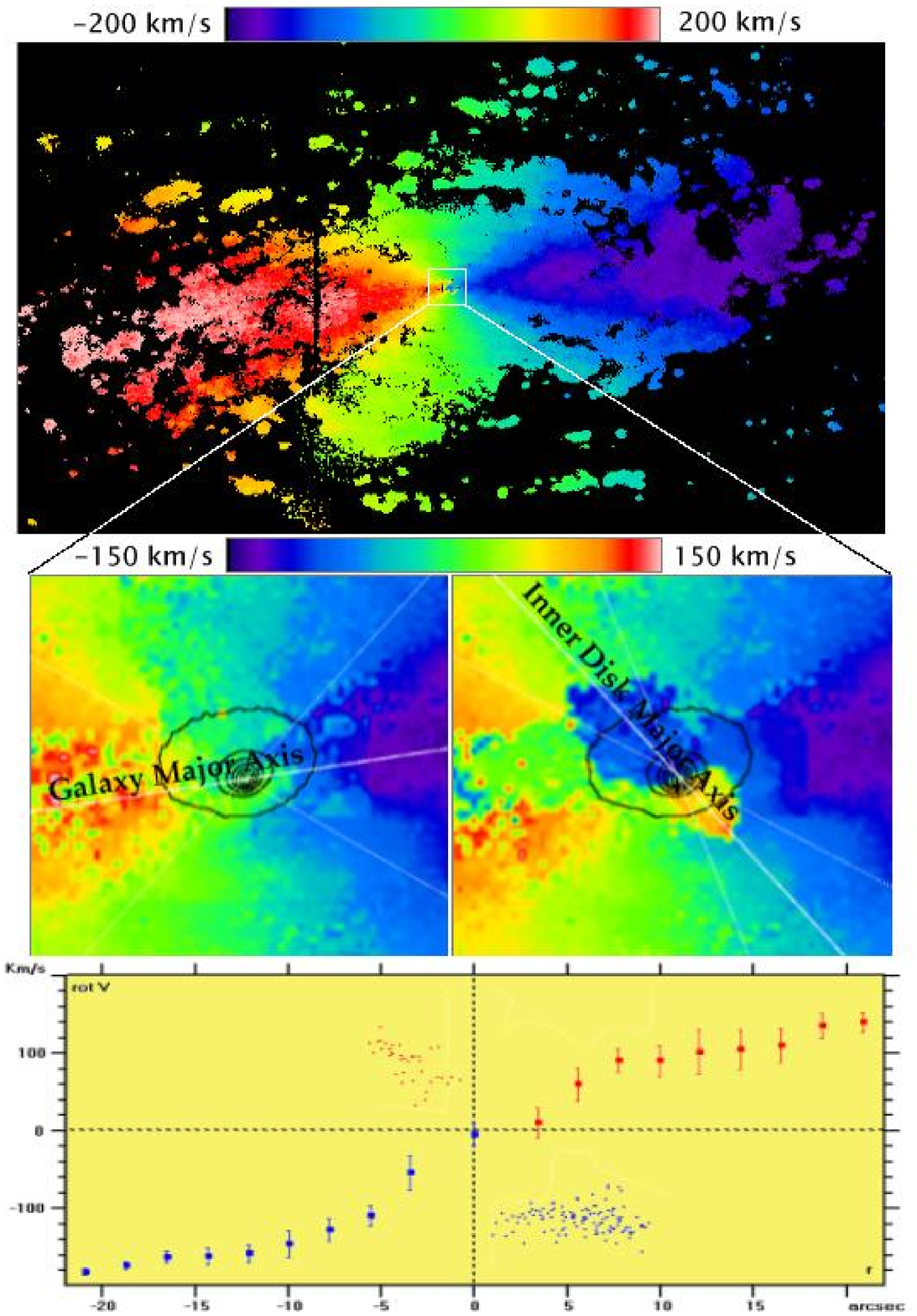}
\caption{{\it Top} Velocity field at full resolution for NGC 5055. {\it Middle} 
  enlargements of the central regions for the two velocity components.
  {\it Bottom} velocity curve (4 pixels bins) for the main disk
  component and radial velocity minus systemic velocity (individual
  pixels) for the central outflow/disk.
\label{fig:5055_BH}}
\end{figure*}

The $\sim 8$ arcsec region around the center of this galaxy is very
peculiar and will be termed "the central region".  At the adopted
distance of 9.2 Mpc, the linear extension is $\pm 360$ pc.  This dense
central region is roughly 6 times brighter in the continuum than the
nearby surrounding region and about 25 times in the H$\alpha$ line.
An important fraction of the pixels of this central region shows two
emission lines. When separated in two velocity components (fig.
\ref{fig:5055_BH}) two drastically different velocity patterns appear.
One pattern shows slightly disturbed kinematics, but still consistent
with the global disk kinematics. The second component shows two
regions of high peculiar velocities compatible with an almost
counter-rotating disk or with a bipolar outflow. In case of a disk,
its position angle is 220\degr, 121\degr~from the galaxy major axis.
The disk inclination can hardly be retrieve due to the small region
with clear nuclear emission line.  Peculiar spectral features seems to
affect a region extending to around 18\arcsec, but they are mostly
buried in the main disk component. Because of this extent,
inclination is most probably lower than 70\degr, and V$\sin i \simeq
111$ \kms up to 300 pc. The systemic velocity is 10\kms lower than the
galaxy systemic velocity, a difference similar to the velocity error
bars.  These velocities lead to a M$\sin i \simeq 9\pm2 \times 10^8$
\msol inside 300 pc.

However, the quasi constant velocities throughout this component, and
the much brighter blue side, are more consistent with a bipolar
outflow.  \cite{Pismis95} already interpreted a departure from pure
rotation on the blue side as such. Their long slit data did not allow
the detection of the red component, but they probably detected the
bright blue clump north-east of the center. However, \cite{Afanasiev02}
note that a south-west anomaly (our red clump) is also present in
their stellar velocities, a rather puzzling fact if the outflow
interpretation is retained. 

\subsection{NGC 2841}

\begin{figure*}[htb]
  \includegraphics[angle=0,width=\textwidth]{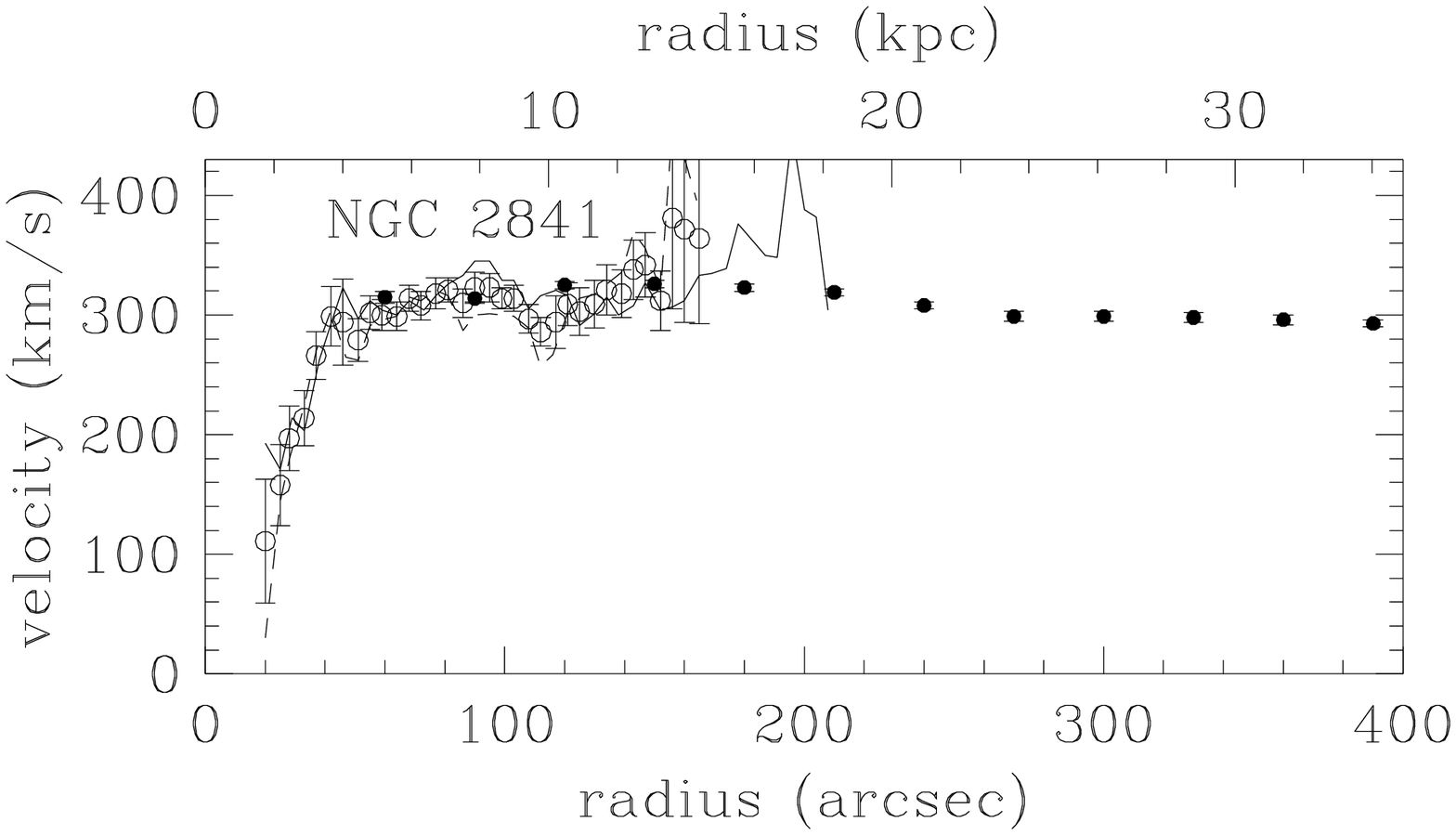}
\caption{\ha rotation curve of NGC 2841 (open circles). Approaching
and receding sides from \ha data are indicated respectively by the
dashed and continuous lines. The filled circles represent the beam
smearing corrected \hI rotation curve from \cite{BegemanPhd}.
\label{fig:rc_2841}}
\end{figure*}

First observed at CFHT with the same blueshifted filter as NGC 5055,
this galaxy was reobserved in December 2001 at the Observatoire de
Haute-Provence (France) in the context of the GHASP survey.  The
instrumentation used at the OHP is very similar to the one used at the
CFHT with the exception that the detector was a photon counting system
instead of a CCD. Because photon counting cameras have intrinsically
no read-out noise, it is possible to rapidly scan through the 24
channels several times, averaging photometric variations of the sky.
It is then easier to remove night sky emission lines and background
light \citep{Amram95}. The filter used at the OHP was matched to the
receding side of the galaxy. Joining the two velocity fields using
stars in the field, a complete coverage was achieved. Despite filters
well matched to the galaxy velocities, we confirm the overall low \ha
luminosity already noted by \cite{Kennicutt88}.  NGC 2841 is one of
the coldest galaxies in the entire IRAS survey having a 60/100 ratio
as low as M31, another galaxy greatly deficient in H-alpha. Data from
2MASS show normal effective colors with J-H = 0.71 and H-K = 0.23
implying a normal visible extinction, thus an intrinsic weak \ha
emission.

Using either a tilted ring model or by matching the two sides of the
galaxy with radially constant values of inclination and position
angle, we found an inclination of 66\degr, a position angle of
150\degr~, and a systemic velocity of 633\kms, virtually identical to
the values found by Begeman inside a radius of 180\arcsec. The
kinematical center was found to be centered on the bright nucleus. The
final \ha curve (Tab. \ref{fig:rc_2841} and Tab. \ref{tab:rc_2841_ha})
is based on a Gaussian smoothed ($\sigma =$ 3\arcsec ) version of the
velocity field.

Despite the very similar orientation parameters, the \ha velocities
stay up to 26\kms below the beam smearing corrected \hI data in the
inner part of the galaxy.  Due to the low radio resolution
($>$32\arcsec) and the steepness of the rotation curve, it is
conceivable that the first \hI points have been over-corrected for
beam smearing. 

\subsection{NGC 5985}

\begin{figure*}[htb]
\includegraphics[angle=0,width=\textwidth]{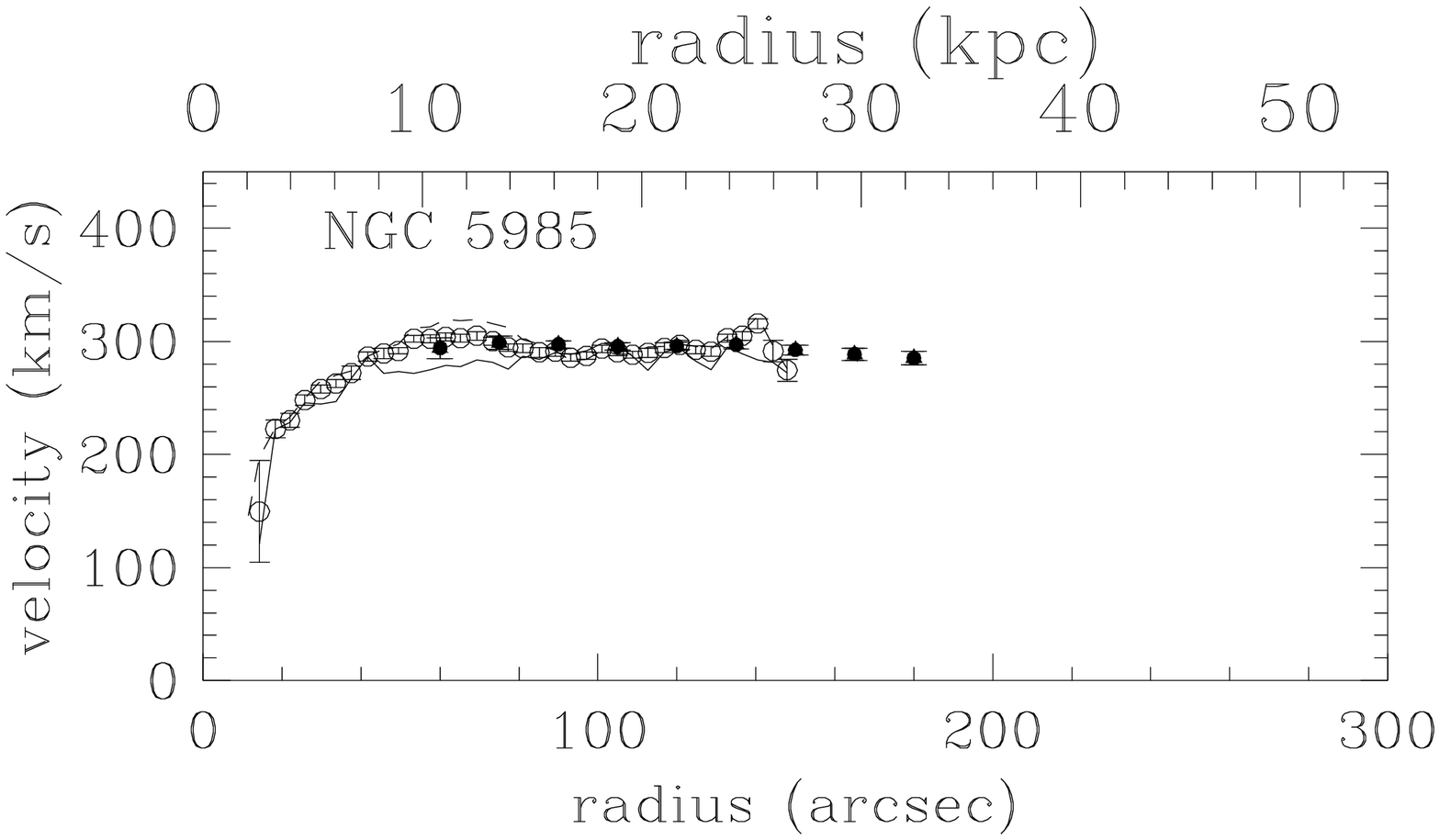}
\caption{\ha rotation curve of NGC 5985 (open circles). Approaching and
receding sides from \ha data are indicated respectively by the dashed
and continuous lines. The filled circles represent the \hI rotation
curve corrected for beam smearing.
\label{fig:rc_5985}}
\end{figure*}

This massive galaxy was observed with a wider filter (19\AA) but
centered 3\AA~bluer than the galaxy central wavelength. That, in
conjunction with the high rotational velocity, prevented the reddest
flux to reach the CCD. The velocity field is thus slightly incomplete
on the receding side but a complete and reliable rotation curve can be
derived using data away from the major axis. There is therefore no
cut-off in rotation velocity but instead, an added uncertainty
throughout the velocity plateau due to the lack of purely radially
moving gas. The \ha rotation curve was retrieved matching the two sides
of the galaxy. The inclination and position angles were found to be
respectively 58\degr~and 15\degr.

For this galaxy an HI rotation curve has been derived from the
WHISP\footnote{www.astro.rug.nl/$\sim$whisp} data. The general
procedures for the observations and reduction of the WHISP data are
presented in \cite{Swaters00}, but details for this particular
galaxy will be published in a future paper. The velocity field for
NGC~5985 was derived by fitting Gaussians to the line profiles. Then,
from a tilted-ring fit to the velocity field derived from the HI
observations we found that the inclination varies little and ranges
between 62\degr~and 58\degr, and that the rotatio angle is nearly
constant at 15\degr.  We therefore fixed the inclination at
60\degr~and the position angle at 15\degr, very close to their \ha
counterparts.

Note that because of the lack of \hI in the center, no \hI rotation curve
can be derived there. This shows the complementarity between \ha and
\hI observations. Not only the former have higher resolution, thus
reducing the problems associated with beam smearing, but also it may
trace the rotation curve in regions where no \hI has been detected.

\section{Summary and Conclusions}                                    
\label{sec:discussion}  

It is well known that \hI observations are commonly affected by beam
smearing. The present data set is well suited to illustrate this fact.
A useful test to estimate the reliability of \hI observations is to
compare the size of the radio beam width to the size of the galaxy.
\cite{BosmaPhd} introduced the ratio $R_B \equiv R_H$/S where $R_H$ is
the Holmberg radius and S is the radio beam major axis width. A more
surface brightness independent ratio was introduced by
\cite{vdBosch00}: $R_{vdB} \equiv \alpha^{-1}$/S where $\alpha^{-1}$
is the B--band stellar disk scale length. Table \ref{tab:bs} compares
those two ratios to the maximum difference between the \ha and the \hI
on the rising part of the rotation curves.

NGC 5055, which shows the largest $R_B$ and $R_{vdB}$, also has the
largest $\Delta V_{max}$. Looking at Fig. \ref{fig:rc_5055}, the
effect of beam smearing seems clearly visible in the inner 2 kpc.  For
the galaxy with the second largest $R_B$, NGC 2403, the beam smearing
corrections of \cite{BegemanPhd} seem to work fine since there is a
very good agreement ($\Delta V_{max} \leq 6.5$ \kms) between the \hI
and the \ha observations. However, for NGC 2841, it appears that the
\hI velocities were overcorrected for beam smearing by
\cite{BegemanPhd}. The results for those two galaxies show the
difficulties related to beam smearing correction, and that high
resolution data are still the best way to get the proper kinematics in
the inner parts.

Not much can be said on the possible beam smearing effects for the 3
other galaxies since UGC 2259 has no \ha in the center, NGC 5985 has
no \hI and NGC 6946 has a too high velocity dispersion which could be
due to the presence of a bar.  As was pointed out by \cite{Swaters99},
the magnitude of beam smearing does not only depends on the size of
the beam, but also on details of the galaxy, such as its inclination,
HI distribution and kinematics.

\begin{table}
\caption{Beam--to--galaxy size and beam smearing importance}
\begin{minipage}{\hsize}
\renewcommand{\footnoterule}{\rule{0pt}{0pt} \vspace{-2mm}}
\label{tab:bs}
\begin{tabular}{lcccc}\hline \hline
Galaxy name & Beam width (S) & $R_B$ & $R_{vdB}$ & $\Delta V_{max}$ \\
&\arcsec& & \kms \\ \hline
UGC 2259 & 13.2 & 5.2 & 1.29 & 7.8   \\
NGC 2403 & 26 & 26 & 4.62 & 6.5   \\
NGC 6946 & 24.5 & 19 & 4.70 & 16.8  \\
NGC 5055 & 12.8 & 46 & 8.51 & 23.9  \\
NGC 2841\footnote{\hI corrected for beam smearing}& 25 & 15 & 2.08 & -34.0 \\
NGC 5985 & 30 & 5.3 & 1 & $\ge$10\footnote{No \hI data for the inner part} \\ \hline
\end{tabular}
\end{minipage}
\end{table}

Amid the uncertainties that can affect rotation curves, non-circular
motions like bars and local phenomena such as expanding bubbles or
spiral arms can be a major source of error. Classical 1-D spectroscopy
at any wavelength, while reaching a sufficient resolution, cannot
easily deal with these problems. On the other hand, these phenomena
can be efficiently tracked by the 2-D spectroscopic coverage from
Fabry-Perot data  \citep[see also][for a similar conclusion
in the context of distance indicators]{Schommer93}. Depending on their
orientation, bars can change the radial velocities in the inner parts
of galaxies. Their signature may however be seen in the velocity field
as a regular deviation from the typical projected circular velocities.
None of the SB or SAB in our sample shows signs of these deviations though
some of them have very low signal to noise in the center (UGC 2259 and
NGC 5985) or very high velocity dispersion (NGC 6946). NGC 5055 shows
strong activity inside 300 pc which can be interpreted either has an
almost counter-rotating disk hosting a $9\pm2 \times 10^8$ \msol
object, or as a bipolar outflow. Gas kinematics alone tend to favor
the bipolar outflow but peculiar stellar velocities in
\cite{Afanasiev02} data would be more consistent with a rotating disk.

Complete 2-D velocity fields also allow independent determinations of
inclination and position angles, kinematical centers and systemic
velocities. This is a major gain over 1-D spectroscopy
considering the sensitivity of the rotation curves to these
parameters.

\begin{acknowledgements} 
  
  We warmly thank Jacques Boulesteix for his software and for fruitful
  discussions on Fabry-Perot data reduction, and the staff of the CFHT
  for their support during the Fabry-Perot data acquisition.  We are indebted
  to St\'ephanie C\^ot\'e for the opportunity to re-observe NGC 5055
  and to Olivier Hernandez who helped with the reduction of the NGC
  2841 data.  CC acknowledges grants from NSERC (Canada) and FQRNT
  (Qu\'ebec).
  
\end{acknowledgements} 


\bibliography{ref}

\begin{thebibliography}{}

\bibitem[\protect\astroncite{{Afanasiev} and {Sil'chenko}}{2002}]{Afanasiev02}
{Afanasiev}, V.~L. and {Sil'chenko}, O.~K., 2002,
\newblock {\aap} {388}, 461

\bibitem[\protect\astroncite{{Amram} et~al.}{1996}]{Amram96}
{Amram}, P., {Balkowski}, C., {Boulesteix}, J., {Cayatte}, V., {Marcelin}, M.,
  and {Sullivan}, III, W., 1996,
\newblock {A\&A} {310}, 737

\bibitem[\protect\astroncite{{Amram} et~al.}{1995}]{Amram95}
{Amram}, P., {Boulesteix}, J., {Marcelin}, M., {Balkowski}, C., {Cayatte}, V.,
  and {Sullivan}, W., 1995,
\newblock {A\&AS} {113}, 35

\bibitem[\protect\astroncite{{Beauvais} and {Bothun}}{2001}]{Beauvais01}
{Beauvais}, C. and {Bothun}, G., 2001,
\newblock {\apjss} {136}, 41

\bibitem[\protect\astroncite{{Begeman}}{1987}]{BegemanPhd}
{Begeman}, K., 1987,
\newblock {Ph.D. thesis}, Rijksuniversiteit Groningen

\bibitem[\protect\astroncite{{Blais-Ouellette} et~al.}{2001}]{paperII}
{Blais-Ouellette}, S.~., {Amram}, P., and {Carignan}, C., 2001,
\newblock {\aj} {121}, 1952,
\newblock (Paper II)

\bibitem[\protect\astroncite{{Blais-Ouellette} et~al.}{1999}]{paperI}
{Blais-Ouellette}, S., {Carignan}, C., {Amram}, P., and {C\^ot\'e}, S., 1999,
\newblock {\aj} {118}, 2123,
\newblock (Paper I)

\bibitem[\protect\astroncite{{Bonnarel} et~al.}{1988}]{Bonnarel88}
{Bonnarel}, F., {Boulesteix}, J., {Georgelin}, Y.~P., {Lecoarer}, E.,
  {Marcelin}, M., {Bacon}, R., and {Monnet}, G., 1988,
\newblock {\aap} {189}, 59

\bibitem[\protect\astroncite{{Bosma}}{1978}]{BosmaPhd}
{Bosma}, A., 1978,
\newblock {Ph.D. thesis}, Rijksuniversiteit Groningen

\bibitem[\protect\astroncite{{Carignan} et~al.}{1990}]{Car90}
{Carignan}, C., {Charbonneau}, P., {Boulanger}, F., and {Viallefond}, F., 1990,
\newblock {A\&A} {234}, 43

\bibitem[\protect\astroncite{{Carignan} et~al.}{1988}]{Car88}
{Carignan}, C., {Sancisi}, R., and {Van Albada}, T.~S., 1988,
\newblock {AJ} {95}, 37

\bibitem[\protect\astroncite{{C\^ot\'e} et~al.}{1991}]{Cote91}
{C\^ot\'e}, S., {Carignan}, C., and {Sancisi}, R., 1991,
\newblock {\aj} {102}, 904

\bibitem[\protect\astroncite{{de Blok} et~al.}{2003}]{deBlok03}
{de Blok}, W.~J.~G., {Bosma}, A., and {McGaugh}, S., 2003,
\newblock {\mnras} {340}, 657

\bibitem[\protect\astroncite{{de Vaucouleurs} and
  {Freeman}}{1972}]{deVaucouleurs72}
{de Vaucouleurs}, G. and {Freeman}, K.~C., 1972,
\newblock {Vistas in Astronomy} {14}, 163

\bibitem[\protect\astroncite{{Freedman}}{1990}]{Freedman90}
{Freedman}, W.~L., 1990,
\newblock {\apj} {355}, L35

\bibitem[\protect\astroncite{{Garrido} et~al.}{2002}]{Garrido02}
{Garrido}, O., {Marcelin}, M., {Amram}, P., and {Boulesteix}, J., 2002,
\newblock {\aap} {387}, 821

\bibitem[\protect\astroncite{{Gonzalez-Serrano} and
  {Valentijn}}{1991}]{Gonzalez91}
{Gonzalez-Serrano}, J.~I. and {Valentijn}, E.~A., 1991,
\newblock {\aap} {242}, 334

\bibitem[\protect\astroncite{{Israel} and {Baas}}{2001}]{Israel01}
{Israel}, F.~P. and {Baas}, F., 2001,
\newblock {\aap} {371}, 433

\bibitem[\protect\astroncite{{Kennicutt}}{1988}]{Kennicutt88}
{Kennicutt}, R.~C., 1988,
\newblock {\apj} {334}, 144

\bibitem[\protect\astroncite{{Macri} et~al.}{2001}]{Macri01}
{Macri}, L.~M., {Stetson}, P.~B., {Bothun}, G.~D., {Freedman}, W.~L.,
  {Garnavich}, P.~M., {Jha}, S., {Madore}, B.~F., and {Richmond}, M.~W., 2001,
\newblock {\apj} {559}, 243

\bibitem[\protect\astroncite{{Moore} et~al.}{1999}]{Moore99}
{Moore}, B., {Ghigna}, S., {Governato}, F., {Lake}, G., {Quinn}, T., {Stadel},
  J., and {Tozzi}, P., 1999,
\newblock {\apjl} {524}, L19

\bibitem[\protect\astroncite{{Moore} et~al.}{1998}]{Moore98}
{Moore}, B., {Governato}, F., {Quinn}, T., {Stadel}, J., and {Lake}, G., 1998,
\newblock {\apj} {499}, L5

\bibitem[\protect\astroncite{{Navarro} et~al.}{1996}]{NFW96}
{Navarro}, J.~F., {Frenk}, C.~S., and {White}, S. D.~M., 1996,
\newblock {ApJ} {462}, 563

\bibitem[\protect\astroncite{{Navarro} et~al.}{1997}]{NFW97}
{Navarro}, J.~F., {Frenk}, C.~S., and {White}, S. D.~M., 1997,
\newblock {ApJ} {490}, 493

\bibitem[\protect\astroncite{{Navarro} et~al.}{2003}]{Navarro03}
{Navarro}, J.~F., {Hayashi}, E., {Power}, C., {Jenkins}, A., {Frenk}, C.,
  {White}, S., {Springel}, V., {Stadel}, J., and {Quinn}, T., 2003,
\newblock {The Inner Structure of LambdaCDM Halos III: Universality and
  Asymptotic Slopes},
\newblock astro-ph/0311231

\bibitem[\protect\astroncite{{Pismis} et~al.}{1995}]{Pismis95}
{Pismis}, P., {Mampaso}, A., {Manteiga}, M., {Recillas}, E., and {Cruz
  Gonzalez}, G., 1995,
\newblock {\aj} {109}, 140

\bibitem[\protect\astroncite{{Roussel} et~al.}{2001}]{Roussel01}
{Roussel}, H., {Vigroux}, L., {Bosma}, A., {Sauvage}, M., {Bonoli}, C.,
  {Gallais}, P., {Hawarden}, T., {Lequeux}, J., {Madden}, S., and {Mazzei}, P.,
  2001,
\newblock {\aap} {369}, 473

\bibitem[\protect\astroncite{{Schommer} et~al.}{1993}]{Schommer93}
{Schommer}, R.~A., {Bothun}, G.~D., {Williams}, T.~B., and {Mould}, J.~R.,
  1993,
\newblock {\aj} {105}, 97

\bibitem[\protect\astroncite{{Sharina} et~al.}{1997}]{Sharina97}
{Sharina}, M.~E., {Karachentsev}, I.~D., and {Tikhonov}, N.~A., 1997,
\newblock {Pis ma Astronomicheskii Zhurnal} {23}, 430

\bibitem[\protect\astroncite{{Sofue}}{1996}]{Sofue96}
{Sofue}, Y., 1996,
\newblock {\apj} {458}, 120

\bibitem[\protect\astroncite{{Sofue} et~al.}{1999}]{Sofue99}
{Sofue}, Y., {Tutui}, Y., {Honma}, M., {Tomita}, A., {Takamiya}, T., {Koda},
  J., and {Takeda}, Y., 1999,
\newblock {\apj} {523}, 136

\bibitem[\protect\astroncite{{Swaters}}{1999a}]{Swaters99}
{Swaters}, R., 1999a,
\newblock in {ASP Conf. Ser. 182: Galaxy Dynamics - A Rutgers Symposium}, p.
  369

\bibitem[\protect\astroncite{{Swaters}}{1999b}]{SwatersPhd}
{Swaters}, R., 1999b,
\newblock {Ph.D. thesis}, Rijksuniversiteit Groningen

\bibitem[\protect\astroncite{{Swaters} et~al.}{2000}]{Swaters00}
{Swaters}, R.~A., {Madore}, B.~F., and {Trewhella}, M., 2000,
\newblock {\apjl} {531}, L107

\bibitem[\protect\astroncite{{Swaters} et~al.}{2003}]{Swaters03}
{Swaters}, R.~A., {Madore}, B.~F., {van den Bosch}, F.~C., and {Balcells}, M.,
  2003,
\newblock {\apj} {583}, 732

\bibitem[\protect\astroncite{{Thornley} and {Mundy}}{1997}]{Thornley97}
{Thornley}, M.~D. and {Mundy}, L.~G., 1997,
\newblock {ApJ} {484}, 202

\bibitem[\protect\astroncite{{van den Bosch} et~al.}{2000}]{vdBosch00}
{van den Bosch}, F.~C., {Robertson}, B.~E., {Dalcanton}, J.~J., and {de Blok},
  W. J.~G., 2000,
\newblock {\aj} {119}, 1579

\bibitem[\protect\astroncite{{van der Kruit} and {Bosma}}{1978}]{vanDerKruit78}
{van der Kruit}, P.~C. and {Bosma}, A., 1978,
\newblock {\aaps} {34}, 259

\bibitem[\protect\astroncite{{Weldrake} et~al.}{2003}]{Weldrake03}
{Weldrake}, D.~T.~F., {de Blok}, W.~J.~G., and {Walter}, F., 2003,
\newblock {\mnras} {340}, 12

\end{thebibliography}



\onecolumn
\appendix
\section{Rotation Curves}
\label{rc_tables}
\bigskip

{\small
\tablecaption{Optical rotation curve of UGC 2259 at 2.6\arcsec\, resolution.  
\label{tab:rc_2259_ha}}
\tablehead{
\multicolumn{6}{l}{{\bf Table~\ref{tab:rc_2259_ha}} {\sl continued}}\\
\noalign{\smallskip}
\hline
\noalign{\smallskip}
R & N$_{app}$ & V$_{app}$ & $\sigma_{ring}$ & N$_{rec}$ & V$_{rec}$ & $\sigma_{ring}$ & V \\  
arcsec && \kms & \kms & & \kms & \kms & \kms  \\ 
\noalign{\smallskip}
\hline
\noalign{\smallskip}
}
\tablefirsthead{
\hline
\noalign{\smallskip}
R & N$_{app}$ & V$_{app}$ & $\sigma_{ring}$ & N$_{rec}$ & V$_{rec}$ & $\sigma_{ring}$ & V \\  
arcsec && \kms & \kms & & \kms & \kms & \kms  \\
\noalign{\smallskip}
\hline
\noalign{\smallskip}
}
\tabletail{
\noalign{\smallskip}
\hline
}
\begin{supertabular}{llllllll}
1.4     &       &       &        & 5    & 0     & 28    & 0     $\pm$ 12        \\
3.8     &       &       &        & 15   & 16    & 9     & 16    $\pm$ 2         \\
6.4     &       &       &        & 17   & 28    & 13    & 28    $\pm$ 3         \\
9.3     &       &       &        & 7    & 35    & 9     & 35    $\pm$ 3         \\
11.6    &       &       &        & 15   & 45    & 15    & 45    $\pm$ 4         \\      
13.7    &       &       &        & 2    & 80    & 8     & 80    $\pm$ 5         \\
17.4    & 14    & 55    & 5      &      &       &       & 55    $\pm$ 1         \\
19.8    & 18    & 63    & 8      &      &       &       & 63    $\pm$ 2         \\
22.8    & 16    & 64    & 10     &      &       &       & 64    $\pm$ 2         \\
24.9    & 14    & 66    & 10     &      &       &       & 66    $\pm$ 3         \\
28.0    & 41    & 68    & 12     &      &       &       & 68    $\pm$ 2         \\
30.4    & 79    & 66    & 13     &      &       &       & 66    $\pm$ 1         \\
33.2    & 80    & 74    & 14     & 6    & 73    & 10    & 74    $\pm$ 2         \\      
35.6    & 51    & 73    & 11     & 23   & 72    & 8     & 72    $\pm$ 2         \\
38.4    & 57    & 78    & 12     & 53   & 72    & 6     & 75    $\pm$ 1         \\
40.9    & 72    & 79    & 14     & 59   & 78    & 8     & 79    $\pm$ 1         \\
43.6    & 92    & 80    & 13     & 74   & 80    & 9     & 80    $\pm$ 1         \\
46.2    & 84    & 82    & 11     & 73   & 82    & 8     & 82    $\pm$ 1         \\
48.9    & 65    & 83    & 9      & 51   & 87    & 5     & 85    $\pm$ 1         \\
51.4    & 46    & 84    & 14     & 38   & 84    & 8     & 84    $\pm$ 2         \\
54.0    & 45    & 86    & 14     & 24   & 79    & 10    & 84    $\pm$ 2         \\
56.7    & 35    & 85    & 10     & 12   & 76    & 9     & 83    $\pm$ 2         \\
59.4    & 25    & 74    & 12     & 13   & 68    & 6     & 72    $\pm$ 2         \\
61.9    & 19    & 71    & 8     & 38    & 69    & 5     & 69    $\pm$ 1         \\
64.3    & 7     & 75    & 5      & 41   & 69    & 6     & 70    $\pm$ 1         \\
67.3    &       &       &        & 28   & 75    & 15    & 75    $\pm$ 3         \\
69.4    &       &       &        & 15   & 76    & 16    & 76    $\pm$ 4         \\
\end{supertabular}

\vfill\eject

\tablecaption{Optical rotation curve of NGC 2403 at 6.6\arcsec\, resolution.  
\label{tab:rc_2403_ha}}
\tablehead{
\multicolumn{6}{l}{{\bf Table~\ref{tab:rc_2403_ha}} {\sl continued}}\\
\noalign{\smallskip}
\hline
\noalign{\smallskip}
R & N$_{app}$ & V$_{app}$ & $\sigma_{ring}$ & N$_{rec}$ & V$_{rec}$ & $\sigma_{ring}$ & V \\  
arcsec && \kms & \kms & & \kms & \kms & \kms  \\
\noalign{\smallskip}
\hline
\noalign{\smallskip}
}
\tablefirsthead{
\hline
\noalign{\smallskip}
R & N$_{app}$ & V$_{app}$ & $\sigma_{ring}$ & N$_{rec}$ & V$_{rec}$ & $\sigma_{ring}$ & V \\  
arcsec && \kms & \kms & & \kms & \kms & \kms  \\
\noalign{\smallskip}
\hline
\noalign{\smallskip}
}
\tabletail{
\noalign{\smallskip}
\hline
}
\begin{supertabular}{llllllll} 
10.3     & 49    & 29     &     13       & 48    & 21     &     9        & 25  $\pm$    2 \\
16.7     & 79    & 41     &     14       & 79    & 32     &     11       & 36  $\pm$    1 \\
23.2     & 116   & 45     &     15       & 116   & 43     &     15       & 44  $\pm$    1 \\
29.96    & 148   & 55     &     16       & 148   & 52     &     14       & 53  $\pm$    1 \\
36.4     & 178   & 67     &     18       & 179   & 56     &     15       & 62  $\pm$    1 \\
43.0     & 209   & 68     &     22       & 209   & 65     &     15       & 67  $\pm$    1 \\
49.6     & 251   & 75     &     23       & 251   & 72     &     16       & 73  $\pm$    1 \\
56.2     & 274   & 78     &     23       & 274   & 73     &     19       & 76  $\pm$    1 \\
62.8     & 312   & 77     &     21       & 312   & 76     &     20       & 76  $\pm$    1 \\
69.4     & 342   & 79     &     19       & 342   & 77     &     18       & 78  $\pm$    1 \\
76.0     & 361   & 83     &     17       & 375   & 78     &     14       & 80  $\pm$    1 \\
83.0     & 379   & 88     &     20       & 408   & 82     &     15       & 85  $\pm$    1 \\
89.1     & 425   & 91     &     19       & 441   & 86     &     15       & 88  $\pm$    1 \\
95.7     & 451   & 91     &     21       & 471   & 87     &     15       & 89  $\pm$    1 \\
102.3    & 479   & 94     &     15       & 506   & 90     &     18       & 92  $\pm$    1 \\
108.9    & 538   & 98     &     21       & 540   & 93     &     19       & 95  $\pm$    1 \\
115.5    & 576   & 97     &     21       & 574   & 93     &     19       & 95  $\pm$    1 \\
122.1    & 600   & 98     &     22       & 583   & 95     &     21       & 97  $\pm$    1 \\
128.7    & 629   & 102    &     19       & 598   & 97     &     22       & 99  $\pm$    1 \\
135.4    & 659   & 104    &     20       & 627   & 97     &     22       & 101 $\pm$    1 \\
141.9    & 695   & 103    &     19       & 651   & 96     &     19       & 100 $\pm$    1 \\
148.5    & 728   & 102    &     20       & 634   & 97     &     18       & 100 $\pm$    1 \\
155.1    & 762   & 100    &     20       & 610   & 100    &     16       & 100 $\pm$    1 \\
161.8    & 788   & 98     &     21       & 667   & 100    &     18       & 99  $\pm$    1 \\
168.3    & 806   & 98     &     20       & 711   & 102    &     19       & 100 $\pm$    1 \\
174.9    & 852   & 99     &     21       & 766   & 101    &     19       & 100 $\pm$    1 \\
181.5    & 870   & 98     &     20       & 779   & 103    &     18       & 100 $\pm$    1 \\
188.1    & 926   & 97     &     23       & 753   & 107    &     19       & 101 $\pm$    1 \\
194.7    & 918   & 98     &     23       & 690   & 108    &     22       & 102 $\pm$    1 \\
201.3    & 948   & 99     &     28       & 698   & 111    &     22       & 104 $\pm$    1 \\
207.9    & 949   & 98     &     30       & 661   & 111    &     20       & 103 $\pm$    1 \\
214.5    & 1010  & 98     &     30       & 656   & 111    &     20       & 103 $\pm$    1 \\
221.1    & 1069  & 101    &     28       & 644   & 112    &     22       & 105 $\pm$    1 \\
227.6    & 1084  & 106    &     23       & 600   & 115    &     21       & 109 $\pm$    1 \\
234.3    & 1076  & 109    &     24       & 510   & 118    &     22       & 112 $\pm$    1 \\
240.9    & 1150  & 111    &     23       & 462   & 124    &     21       & 114 $\pm$    1 \\
247.4    & 1204  & 112    &     26       & 321   & 132    &     25       & 116 $\pm$    1 \\
254.0    & 1133  & 116    &     28       & 304   & 136    &     21       & 120 $\pm$    1 \\
260.7    & 1089  & 116    &     30       & 291   & 135    &     20       & 120 $\pm$    1 \\
267.2    & 1058  & 115    &     30       & 236   & 128    &     20       & 118 $\pm$    1 \\
273.9    & 1127  & 114    &     31       & 163   & 127    &     24       & 115 $\pm$    1 \\
280.6    & 1130  & 112    &     30       & 158   & 127    &     23       & 114 $\pm$    1 \\
287.1    & 1032  & 110    &     27       & 202   & 118    &     21       & 112 $\pm$    1 \\
293.6    & 1009  & 107    &     24       & 182   & 114    &     18       & 108 $\pm$    1 \\
300.1    & 1085  & 110    &     23       & 208   & 118    &     18       & 111 $\pm$    1 \\
306.9    & 1043  & 113    &     24       & 264   & 121    &     15       & 115 $\pm$    1 \\
313.4    & 935   & 115    &     25       & 240   & 119    &     16       & 115 $\pm$    1 \\
320.1    & 1015  & 118    &     26       & 269   & 118    &     16       & 118 $\pm$    1 \\
326.7    & 1068  & 118    &     24       & 202   & 117    &     12       & 118 $\pm$    1 \\
333.3    & 1076  & 121    &     23       & 232   & 121    &     14       & 121 $\pm$    1 \\
339.8    & 833   & 122    &     25       & 201   & 124    &     12       & 123 $\pm$    1 \\
346.5    & 776   & 121    &     25       & 176   & 123    &     12       & 122 $\pm$    1 \\
353.1    & 853   & 120    &     24       & 159   & 124    &     18       & 120 $\pm$    1 \\
359.7    & 834   & 119    &     20       & 160   & 127    &     11       & 120 $\pm$    1 \\
366.3    & 789   & 119    &     23       & 148   & 128    &     11       & 121 $\pm$    1 \\
372.9    & 691   & 122    &     25       & 141   & 129    &     29       & 123 $\pm$    1 \\
379.5    & 640   & 120    &     25       & 116   & 141    &     10       & 123 $\pm$    1 \\
386.1    & 748   & 119    &     22       & 105   & 149    &     7        & 123 $\pm$    1 \\
392.9    & 725   & 118    &     23       & 142   & 153    &     13       & 124 $\pm$    1 \\
399.4    & 759   & 120    &     21       & 172   & 151    &     12       & 125 $\pm$    1 \\
405.6    & 620   & 122    &     29       & 166   & 145    &     17       & 126 $\pm$    1 \\
412.5    & 492   & 123    &     32       & 127   & 145    &     33       & 128 $\pm$    2 \\
418.9    & 485   & 125    &     29       & 106   & 143    &     33       & 128 $\pm$    2 \\
425.3    & 453   & 128    &     28       & 39    & 156    &     29       & 130 $\pm$    2 \\
433.1    & 307   & 128    &     26       & 11    & 115    &     3        & 127 $\pm$    1 \\
438.8    & 270   & 125    &     40       & 23    & 113    &     5        & 124 $\pm$    2 \\
443.9    & 281   & 118    &     45       & 2     & 109    &     0        & 118 $\pm$    3 \\
452.6    & 310   & 121    &     42       & 31    & 129    &     4        & 122 $\pm$    2 \\
458.5    & 318   & 128    &     29       & 40    & 131    &     4        & 129 $\pm$    2 \\
462.9    &       &        &              & 5     & 133    &     9        & 133 $\pm$    4 \\
465.3    & 304   & 134    &     29       &       &        &              & 134 $\pm$    2 \\
471.8    & 278   & 142    &     30       &       &        &              & 142 $\pm$    2 \\
478.3    & 220   & 147    &     33       &       &        &              & 147 $\pm$    2 \\
484.4    & 107   & 145    &     32       &       &        &              & 145 $\pm$    3 \\
490.2    & 15    & 114    &     16       &       &        &              & 114 $\pm$    4 \\
500.3    & 3     & 116    &     15       &       &        &              & 116 $\pm$    9 \\
505.9    & 66    & 133    &     35       &       &        &              & 133 $\pm$    4 \\
511.5    & 139   & 124    &     34       &       &        &              & 124 $\pm$    3 \\
518.2    & 170   & 128    &     29       &       &        &              & 128 $\pm$    2 \\
524.7    & 173   & 130    &     33       &       &        &              & 130 $\pm$    2 \\
531.2    & 148   & 126    &     34       &       &        &              & 126 $\pm$    3 \\
537.6    & 91    & 118    &     19       &       &        &              & 118 $\pm$    2 \\
544.7    & 96    & 125    &     19       &       &        &              & 125 $\pm$    2 \\
551.1    & 128   & 127    &     22       &       &        &              & 127 $\pm$    2 \\
557.8    & 152   & 134    &     25       &       &        &              & 135 $\pm$    2 \\
564.4    & 144   & 129    &     27       &       &        &              & 129 $\pm$    2 \\
570.7    & 130   & 135    &     28       &       &        &              & 135 $\pm$    2 \\
576.8    & 72    & 125    &     26       &       &        &              & 125 $\pm$    3 \\
583.9    & 18    & 117    &     21       &       &        &              & 117 $\pm$    5 \\
589.9    & 7     & 114    &     21       &       &        &              & 114 $\pm$    8 \\
\end{supertabular}

\vfill\eject

\tablecaption{Optical rotation curve of NGC 6946 at 4.4\arcsec\, resolution.  
\label{tab:rc_6946_ha}}
\tablehead{
\multicolumn{6}{l}{{\bf Table~\ref{tab:rc_6946_ha}} {\sl continued}}\\
\noalign{\smallskip}
\hline
\noalign{\smallskip}
R & N$_{app}$ & V$_{app}$ & $\sigma_{ring}$ & N$_{rec}$ & V$_{rec}$ & $\sigma_{ring}$ & V \\  
arcsec && \kms & \kms & & \kms & \kms & \kms  \\
\noalign{\smallskip}
\hline
\noalign{\smallskip}
}
\tablefirsthead{
\hline
\noalign{\smallskip}
R & N$_{app}$ & V$_{app}$ & $\sigma_{ring}$ & N$_{rec}$ & V$_{rec}$ & $\sigma_{ring}$ & V \\  
arcsec && \kms & \kms & & \kms & \kms & \kms  \\
\noalign{\smallskip}
\hline
\noalign{\smallskip}
}
\tabletail{
\noalign{\smallskip}
\hline
}
\begin{supertabular}{llllllll}
6.9     & 70    & 103    &      28      & 66    & 110    &      71      & 107 $\pm$     7       \\
11.2    & 109   & 124    &      27      & 105   & 150    &      22      & 136 $\pm$     2       \\
15.4    & 156   & 141    &      27      & 113   & 126    &      34      & 135 $\pm$     3       \\
19.7    & 201   & 150    &      30      & 88    & 115    &      47      & 140 $\pm$     3       \\
24.4    & 243   & 139    &      33      & 76    & 85     &      56      & 126 $\pm$     4       \\
28.7    & 247   & 134    &      31      & 119   & 104    &      28      & 124 $\pm$     2       \\
33.0    & 184   & 128    &      39      & 151   & 131    &      41      & 129 $\pm$     3       \\
37.1    & 150   & 108    &      27      & 118   & 134    &      53      & 120 $\pm$     4       \\
41.9    & 155   & 96     &      33      & 71    & 146    &      52      & 112 $\pm$     4       \\
46.4    & 217   & 109    &      44      & 78    & 113    &      48      & 110 $\pm$     4       \\
50.8    & 292   & 108    &      46      & 96    & 97     &      62      & 106 $\pm$     4       \\
55.1    & 390   & 113    &      45      & 221   & 116    &      39      & 114 $\pm$     2       \\
59.5    & 428   & 120    &      38      & 301   & 117    &      28      & 119 $\pm$     2       \\
63.8    & 420   & 118    &      35      & 423   & 131    &      26      & 125 $\pm$     1       \\
68.3    & 449   & 118    &      38      & 421   & 138    &      20      & 128 $\pm$     1       \\
72.6    & 529   & 118    &      24      & 425   & 140    &      19      & 128 $\pm$     1       \\
77.0    & 626   & 117    &      25      & 419   & 144    &      19      & 128 $\pm$     1       \\
81.5    & 707   & 114    &      22      & 579   & 143    &      20      & 127 $\pm$     1       \\
85.8    & 682   & 111    &      23      & 615   & 138    &      22      & 124 $\pm$     1       \\
90.2    & 682   & 111    &      21      & 668   & 139    &      24      & 125 $\pm$     1       \\
94.6    & 652   & 113    &      26      & 674   & 137    &      21      & 125 $\pm$     1       \\
99.0    & 723   & 113    &      29      & 703   & 135    &      19      & 124 $\pm$     1       \\
103.4   & 635   & 111    &      26      & 553   & 135    &      22      & 122 $\pm$     1       \\
107.8   & 609   & 120    &      20      & 522   & 135    &      23      & 127 $\pm$     1       \\
112.2   & 598   & 128    &      18      & 561   & 134    &      22      & 131 $\pm$     1       \\
116.6   & 703   & 134    &      17      & 552   & 136    &      20      & 135 $\pm$     1       \\
121.0   & 661   & 139    &      18      & 649   & 142    &      17      & 141 $\pm$     1       \\
125.5   & 661   & 139    &      20      & 800   & 145    &      16      & 142 $\pm$     1       \\
129.8   & 700   & 143    &      20      & 925   & 148    &      19      & 146 $\pm$     1       \\
134.2   & 680   & 146    &      26      & 1016  & 150    &      29      & 149 $\pm$     1       \\
138.6   & 675   & 137    &      22      & 1052  & 150    &      24      & 145 $\pm$     1       \\
143.0   & 713   & 143    &      18      & 1068  & 148    &      23      & 146 $\pm$     1       \\
147.4   & 668   & 147    &      17      & 1031  & 155    &      21      & 152 $\pm$     1       \\
151.8   & 661   & 152    &      15      & 924   & 157    &      32      & 155 $\pm$     1       \\
156.2   & 782   & 153    &      15      & 719   & 159    &      38      & 156 $\pm$     1       \\
160.7   & 807   & 155    &      14      & 706   & 162    &      36      & 158 $\pm$     1       \\
165.0   & 830   & 156    &      17      & 689   & 161    &      38      & 158 $\pm$     1       \\
169.4   & 863   & 155    &      18      & 684   & 156    &      35      & 155 $\pm$     1       \\
173.8   & 837   & 162    &      16      & 553   & 156    &      37      & 159 $\pm$     1       \\
178.1   & 706   & 164    &      18      & 495   & 160    &      42      & 162 $\pm$     1       \\
182.6   & 533   & 166    &      19      & 399   & 155    &      37      & 161 $\pm$     1       \\
187.0   & 408   & 171    &      21      & 430   & 154    &      30      & 162 $\pm$     1       \\
191.4   & 382   & 174    &      19      & 420   & 158    &      29      & 165 $\pm$     1       \\
195.8   & 396   & 176    &      17      & 309   & 161    &      26      & 169 $\pm$     1       \\
200.2   & 317   & 169    &      15      & 329   & 160    &      35      & 165 $\pm$     2       \\
204.7   & 235   & 171    &      14      & 342   & 164    &      23      & 167 $\pm$     1       \\
209.1   & 295   & 172    &      17      & 346   & 165    &      32      & 168 $\pm$     1       \\
213.4   & 312   & 175    &      16      & 429   & 172    &      20      & 174 $\pm$     1       \\
217.8   & 250   & 174    &      14      & 332   & 175    &      23      & 175 $\pm$     1       \\
222.2   & 193   & 172    &      13      & 504   & 163    &      38      & 165 $\pm$     2       \\
226.7   & 227   & 169    &      16      & 380   & 161    &      23      & 164 $\pm$     1       \\
231.2   & 272   & 180    &      20      & 411   & 160    &      27      & 168 $\pm$     1       \\
235.3   & 286   & 187    &      24      & 607   & 154    &      35      & 165 $\pm$     1       \\
239.9   & 344   & 188    &      27      & 471   & 156    &      26      & 169 $\pm$     1       \\
244.2   & 497   & 182    &      26      & 443   & 167    &      24      & 175 $\pm$     1       \\
248.6   & 572   & 177    &      34      & 542   & 163    &      24      & 170 $\pm$     1       \\
253.0   & 409   & 175    &      35      & 628   & 163    &      24      & 168 $\pm$     1       \\
257.4   & 270   & 178    &      32      & 743   & 168    &      24      & 170 $\pm$     1       \\
262.0   & 140   & 188    &      30      & 816   & 169    &      26      & 172 $\pm$     1       \\
266.2   & 154   & 206    &      44      & 797   & 163    &      31      & 170 $\pm$     2       \\
270.2   & 57    & 208    &      46      & 745   & 162    &      28      & 165 $\pm$     2       \\
274.8   & 19    & 238    &      15      & 675   & 158    &      30      & 160 $\pm$     1       \\
279.3   &       &        &              & 563   & 159    &      29      & 159 $\pm$     1       \\
283.7   &       &        &              & 469   & 162    &      30      & 162 $\pm$     1       \\
288.2   &       &        &              & 380   & 178    &      28      & 178 $\pm$     1       \\
292.4   &       &        &              & 283   & 183    &      30      & 183 $\pm$     2       \\
296.7   &       &        &              & 162   & 175    &      31      & 175 $\pm$     2       \\
300.9   &       &        &              & 48    & 154    &      26      & 154 $\pm$     26      \\
\end{supertabular}

\vfill\eject

\tablecaption{Optical rotation curve of NGC 5055 at 2.2\arcsec\, resolution.  
\label{tab:rc_5055_ha}}
\tablehead{
\multicolumn{6}{l}{{\bf Table~\ref{tab:rc_5055_ha}} {\sl continued}}\\
\noalign{\smallskip}
\hline
\noalign{\smallskip}
R & N$_{app}$ & V$_{app}$ & $\sigma_{ring}$ & N$_{rec}$ & V$_{rec}$ & $\sigma_{ring}$ & V \\  
arcsec && \kms & \kms & & \kms & \kms & \kms  \\
\noalign{\smallskip}
\hline
\noalign{\smallskip}
}
\tablefirsthead{
\hline
\noalign{\smallskip}
R & N$_{app}$ & V$_{app}$ & $\sigma_{ring}$ & N$_{rec}$ & V$_{rec}$ & $\sigma_{ring}$ & V \\  
arcsec && \kms & \kms & & \kms & \kms & \kms  \\
\noalign{\smallskip}
\hline
\noalign{\smallskip}
}
\tabletail{
\hline
}
\begin{supertabular}{llllllll}
3       &       36      &       64      &       55      &       36      &       3       &       36      &       34      $\pm$   11      \\
6       &       59      &       111     &       25      &       59      &       64      &       33      &       88      $\pm$   5       \\
8       &       85      &       130     &       31      &       85      &       86      &       38      &       108     $\pm$   5       \\
10      &       108     &       147     &       33      &       108     &       89      &       42      &       118     $\pm$   5       \\
12      &       131     &       170     &       21      &       124     &       143     &       21      &       156     $\pm$   3       \\
14      &       153     &       169     &       17      &       129     &       142     &       22      &       155     $\pm$   2       \\
17      &       182     &       174     &       11      &       172     &       148     &       21      &       161     $\pm$   2       \\
19      &       201     &       179     &       9       &       201     &       150     &       21      &       164     $\pm$   2       \\
21      &       228     &       184     &       13      &       228     &       149     &       15      &       167     $\pm$   1       \\
23      &       254     &       186     &       19      &       254     &       153     &       14      &       169     $\pm$   1       \\
25      &       277     &       184     &       19      &       275     &       156     &       14      &       170     $\pm$   1       \\
28      &       291     &       183     &       16      &       291     &       159     &       17      &       171     $\pm$   1       \\
30      &       323     &       185     &       15      &       315     &       164     &       18      &       175     $\pm$   1       \\
32      &       347     &       186     &       13      &       339     &       170     &       16      &       178     $\pm$   1       \\
34      &       373     &       185     &       15      &       373     &       173     &       17      &       179     $\pm$   1       \\
36      &       387     &       185     &       16      &       384     &       177     &       13      &       181     $\pm$   1       \\
39      &       423     &       188     &       14      &       423     &       178     &       10      &       183     $\pm$   1       \\
41      &       443     &       191     &       12      &       443     &       180     &       10      &       185     $\pm$   1       \\
43      &       462     &       192     &       11      &       460     &       180     &       11      &       186     $\pm$   1       \\
45      &       487     &       193     &       11      &       486     &       180     &       11      &       187     $\pm$   1       \\
47      &       514     &       196     &       11      &       512     &       182     &       10      &       189     $\pm$   1       \\
50      &       544     &       198     &       11      &       512     &       185     &       10      &       191     $\pm$   1       \\
52      &       558     &       199     &       11      &       512     &       192     &       10      &       195     $\pm$   1       \\
54      &       573     &       199     &       10      &       515     &       198     &       13      &       198     $\pm$   1       \\
56      &       606     &       200     &       11      &       506     &       200     &       12      &       200     $\pm$   1       \\
58      &       631     &       201     &       10      &       560     &       202     &       15      &       201     $\pm$   1       \\
61      &       654     &       202     &       10      &       602     &       203     &       15      &       203     $\pm$   1       \\
63      &       686     &       202     &       9       &       604     &       208     &       13      &       205     $\pm$   1       \\
65      &       695     &       201     &       10      &       641     &       209     &       12      &       205     $\pm$   1       \\
67      &       706     &       202     &       10      &       678     &       209     &       18      &       206     $\pm$   1       \\
69      &       718     &       201     &       11      &       647     &       209     &       16      &       205     $\pm$   1       \\
72      &       725     &       200     &       12      &       535     &       207     &       18      &       204     $\pm$   1       \\
74      &       742     &       197     &       16      &       520     &       208     &       15      &       203     $\pm$   1       \\
76      &       778     &       195     &       19      &       573     &       207     &       17      &       201     $\pm$   1       \\
78      &       782     &       195     &       22      &       611     &       205     &       18      &       200     $\pm$   1       \\
80      &       825     &       195     &       20      &       677     &       202     &       19      &       199     $\pm$   1       \\
83      &       841     &       196     &       18      &       686     &       201     &       20      &       199     $\pm$   1       \\
85      &       863     &       195     &       16      &       714     &       198     &       22      &       197     $\pm$   1       \\
87      &       883     &       195     &       15      &       642     &       198     &       21      &       197     $\pm$   1       \\
89      &       833     &       196     &       16      &       617     &       194     &       27      &       195     $\pm$   1       \\
91      &       826     &       197     &       15      &       600     &       195     &       24      &       196     $\pm$   1       \\
93      &       864     &       197     &       15      &       665     &       191     &       26      &       194     $\pm$   1       \\
96      &       893     &       197     &       14      &       724     &       188     &       30      &       192     $\pm$   1       \\
98      &       910     &       197     &       14      &       631     &       191     &       28      &       194     $\pm$   1       \\
100     &       931     &       198     &       15      &       678     &       189     &       28      &       193     $\pm$   1       \\
102     &       935     &       200     &       16      &       792     &       193     &       23      &       196     $\pm$   1       \\
104     &       902     &       202     &       15      &       885     &       197     &       21      &       199     $\pm$   1       \\
107     &       870     &       204     &       14      &       977     &       195     &       27      &       199     $\pm$   1       \\
109     &       928     &       204     &       13      &       955     &       196     &       25      &       200     $\pm$   1       \\
111     &       935     &       204     &       14      &       896     &       198     &       25      &       201     $\pm$   1       \\
113     &       1013    &       203     &       15      &       851     &       200     &       25      &       201     $\pm$   1       \\
116     &       976     &       204     &       14      &       814     &       199     &       29      &       202     $\pm$   1       \\
118     &       987     &       205     &       13      &       799     &       199     &       33      &       202     $\pm$   1       \\
120     &       1046    &       205     &       12      &       778     &       202     &       30      &       203     $\pm$   1       \\
122     &       1100    &       205     &       13      &       771     &       204     &       31      &       204     $\pm$   1       \\
124     &       1219    &       204     &       12      &       802     &       203     &       34      &       203     $\pm$   1       \\
127     &       1268    &       204     &       12      &       839     &       203     &       34      &       204     $\pm$   1       \\
129     &       1282    &       205     &       11      &       913     &       203     &       34      &       204     $\pm$   1       \\
131     &       1323    &       207     &       11      &       957     &       208     &       29      &       207     $\pm$   1       \\
133     &       1259    &       209     &       12      &       918     &       209     &       27      &       209     $\pm$   1       \\
135     &       1214    &       209     &       12      &       810     &       208     &       29      &       209     $\pm$   1       \\
137     &       1160    &       210     &       12      &       741     &       209     &       26      &       210     $\pm$   1       \\
140     &       1111    &       208     &       12      &       623     &       209     &       27      &       209     $\pm$   1       \\
142     &       1116    &       208     &       12      &       602     &       209     &       30      &       209     $\pm$   1       \\
144     &       1093    &       207     &       12      &       649     &       207     &       33      &       207     $\pm$   1       \\
146     &       1132    &       206     &       12      &       624     &       207     &       34      &       206     $\pm$   1       \\
149     &       1153    &       205     &       13      &       605     &       210     &       28      &       208     $\pm$   1       \\
151     &       1175    &       204     &       13      &       619     &       211     &       22      &       208     $\pm$   1       \\
153     &       1175    &       204     &       15      &       622     &       209     &       24      &       207     $\pm$   1       \\
155     &       1230    &       205     &       16      &       587     &       206     &       28      &       206     $\pm$   1       \\
157     &       1177    &       206     &       17      &       530     &       206     &       30      &       206     $\pm$   1       \\
160     &       1080    &       208     &       16      &       458     &       203     &       34      &       205     $\pm$   2       \\
162     &       980     &       208     &       13      &       467     &       200     &       32      &       204     $\pm$   2       \\
164     &       988     &       209     &       11      &       537     &       204     &       26      &       206     $\pm$   1       \\
166     &       1022    &       211     &       12      &       711     &       205     &       22      &       208     $\pm$   1       \\
168     &       990     &       211     &       12      &       630     &       203     &       22      &       207     $\pm$   1       \\
171     &       1032    &       212     &       13      &       588     &       201     &       22      &       206     $\pm$   1       \\
173     &       1077    &       209     &       14      &       606     &       200     &       24      &       205     $\pm$   1       \\
175     &       1009    &       209     &       12      &       570     &       203     &       28      &       206     $\pm$   1       \\
177     &       1050    &       208     &       11      &       542     &       206     &       21      &       207     $\pm$   1       \\
179     &       1018    &       207     &       11      &       507     &       207     &       21      &       207     $\pm$   1       \\
182     &       1004    &       208     &       13      &       529     &       207     &       20      &       207     $\pm$   1       \\
184     &       960     &       208     &       13      &       593     &       203     &       24      &       206     $\pm$   1       \\
186     &       836     &       209     &       14      &       688     &       202     &       27      &       205     $\pm$   1       \\
188     &       674     &       210     &       14      &       591     &       200     &       30      &       205     $\pm$   1       \\
190     &       608     &       212     &       12      &       569     &       199     &       29      &       206     $\pm$   1       \\
192     &       499     &       213     &       11      &       539     &       201     &       28      &       207     $\pm$   1       \\
195     &       472     &       213     &       11      &       623     &       204     &       26      &       208     $\pm$   1       \\
197     &       415     &       211     &       11      &       714     &       204     &       26      &       207     $\pm$   1       \\
199     &       296     &       208     &       14      &       702     &       207     &       25      &       208     $\pm$   1       \\
201     &       276     &       202     &       27      &       599     &       207     &       26      &       205     $\pm$   2       \\
204     &       321     &       198     &       35      &       619     &       210     &       28      &       204     $\pm$   2       \\
206     &       389     &       202     &       19      &       603     &       217     &       17      &       209     $\pm$   1       \\
208     &       341     &       200     &       13      &       575     &       217     &       13      &       208     $\pm$   1       \\
210     &       327     &       199     &       11      &       436     &       215     &       15      &       207     $\pm$   1       \\
212     &       276     &       201     &       17      &       309     &       216     &       17      &       209     $\pm$   1       \\
214     &       193     &       202     &       14      &       283     &       214     &       20      &       208     $\pm$   2       \\
217     &       112     &       199     &       15      &       283     &       208     &       29      &       203     $\pm$   2       \\
219     &       82      &       192     &       28      &       278     &       208     &       27      &       200     $\pm$   3       \\
221     &       67      &       196     &       15      &       252     &       206     &       27      &       201     $\pm$   2       \\
223     &       56      &       208     &       10      &       293     &       213     &       22      &       210     $\pm$   2       \\
226     &       20      &       218     &       19      &       344     &       213     &       19      &       215     $\pm$   4       \\
227     &       10      &       223     &       28      &       376     &       211     &       19      &       217     $\pm$   9       \\
233     &       13      &       190     &       23      &       267     &       208     &       14      &       199     $\pm$   6       \\
234     &       21      &       167     &       17      &       275     &       211     &       19      &       189     $\pm$   4       \\
237     &       29      &       181     &       45      &       282     &       217     &       27      &       199     $\pm$   8       \\
239     &       21      &       174     &       78      &       209     &       217     &       37      &       195     $\pm$   17      \\
241     &       42      &       180     &       20      &       218     &       212     &       33      &       196     $\pm$   4       \\
243     &       29      &       191     &       13      &       138     &       219     &       29      &       205     $\pm$   3       \\
245     &       51      &       192     &       5       &       69      &       221     &       34      &       206     $\pm$   4       \\
247     &       28      &       192     &       7       &       65      &       235     &       52      &       213     $\pm$   7       \\
251     &       1       &       214     &       0       &       96      &       234     &       20      &       224     $\pm$   2       \\
255     &       8       &       204     &       4       &       86      &       237     &       11      &       221     $\pm$   2       \\
256     &       23      &       208     &       17      &       119     &       245     &       11      &       227     $\pm$   4       \\
259     &       43      &       199     &       31      &       165     &       238     &       14      &       219     $\pm$   5       \\
261     &       49      &       208     &       14      &       111     &       238     &       15      &       223     $\pm$   2       \\
263     &       73      &       205     &       17      &       44      &       233     &       21      &       219     $\pm$   4       \\
265     &       78      &       206     &       18      &       33      &       249     &       34      &       227     $\pm$   6       \\
267     &       81      &       204     &       30      &       9       &       218     &       92      &       211     $\pm$   31      \\
270     &       110     &       207     &       21      &       6       &       241     &       15      &       224     $\pm$   6       \\
272     &       133     &       206     &       22      &       16      &       234     &       15      &       220     $\pm$   4       \\
274     &       126     &       200     &       28      &       28      &       231     &       20      &       215     $\pm$   4       \\
276     &       94      &       205     &       21      &       26      &       223     &       14      &       214     $\pm$   3       \\
278     &       92      &       212     &       17      &       21      &       230     &       10      &       221     $\pm$   3       \\
281     &       120     &       217     &       21      &       2       &       139     &       19      &       178     $\pm$   14      \\
283     &       134     &       205     &       25      &       3       &       47      &       4       &       126     $\pm$   3       \\
285     &       128     &       201     &       28      &       1       &       29      &       0       &       115     $\pm$   2       \\
287     &       144     &       196     &       31      &       4       &       195     &       51      &       195     $\pm$   25      \\
289     &       146     &       198     &       30      &       12      &       112     &       53      &       155     $\pm$   15      \\
291     &       112     &       199     &       23      &       7       &       117     &       16      &       158     $\pm$   7       \\
294     &       66      &       199     &       23      &               &               &               &       199     $\pm$   3       \\
296     &       56      &       182     &       24      &               &               &               &       182     $\pm$   3       \\
298     &       86      &       176     &       18      &               &               &               &       176     $\pm$   2       \\
300     &       93      &       177     &       19      &               &               &               &       177     $\pm$   2       \\
303     &       101     &       172     &       23      &               &               &               &       172     $\pm$   2       \\
305     &       118     &       164     &       35      &               &               &               &       164     $\pm$   3       \\
307     &       76      &       160     &       14      &               &               &               &       160     $\pm$   2       \\
\end{supertabular}        

\vfill\eject

\tablecaption{Optical rotation curve of NGC 2841 at 4.4\arcsec\,
resolution. 
\label{tab:rc_2841_ha}}
\tablehead{
\multicolumn{4}{l}{{\bf Table~\ref{tab:rc_2841_ha}} {\sl continued}}\\
\noalign{\smallskip}
\hline
\noalign{\smallskip}
R & N$_{app}$ & V$_{app}$ & $\sigma_{ring}$ & N$_{rec}$ & V$_{rec}$ & $\sigma_{ring}$ & V \\  
arcsec && \kms & \kms & & \kms & \kms & \kms  \\
\noalign{\smallskip}
\hline
\noalign{\smallskip}
}
\tablefirsthead{
\hline
\noalign{\smallskip}
R & N$_{app}$ & V$_{app}$ & $\sigma_{ring}$ & N$_{rec}$ & V$_{rec}$ & $\sigma_{ring}$ & V \\  
arcsec && \kms & \kms & & \kms & \kms & \kms  \\
\noalign{\smallskip}
\hline
\noalign{\smallskip}
}
\tabletail{
\noalign{\smallskip}
\hline
}
\begin{supertabular}{llllllll}
20      &       41      &       193     &       160     &       26      &       30      &       232     &       111     $\pm$   52      \\
25      &       19      &       171     &       103     &       42      &       144     &       155     &       158     $\pm$   34      \\
28      &       21      &       214     &       90      &       46      &       180     &       126     &       197     $\pm$   27      \\
33      &       54      &       203     &       80      &       43      &       225     &       133     &       214     $\pm$   23      \\
37      &       47      &       262     &       112     &       33      &       270     &       62      &       266     $\pm$   20      \\
42      &       58      &       293     &       155     &       30      &       304     &       75      &       299     $\pm$   25      \\
46      &       60      &       323     &       177     &       63      &       265     &       225     &       294     $\pm$   36      \\
51      &       113     &       297     &       161     &       128     &       262     &       117     &       279     $\pm$   18      \\
55      &       151     &       312     &       154     &       231     &       291     &       90      &       302     $\pm$   14      \\
59      &       171     &       304     &       150     &       276     &       296     &       107     &       300     $\pm$   13      \\
64      &       176     &       300     &       142     &       282     &       299     &       89      &       299     $\pm$   12      \\
68      &       237     &       312     &       152     &       227     &       315     &       78      &       314     $\pm$   11      \\
72      &       265     &       304     &       159     &       209     &       312     &       98      &       308     $\pm$   12      \\
77      &       286     &       316     &       153     &       167     &       321     &       113     &       318     $\pm$   13      \\
81      &       344     &       326     &       145     &       130     &       316     &       68      &       321     $\pm$   10      \\
86      &       314     &       333     &       147     &       138     &       287     &       94      &       310     $\pm$   12      \\
90      &       306     &       345     &       159     &       197     &       300     &       129     &       323     $\pm$   13      \\
95      &       311     &       345     &       185     &       240     &       301     &       104     &       323     $\pm$   12      \\
99      &       339     &       329     &       147     &       262     &       300     &       68      &       314     $\pm$   9       \\
103     &       352     &       329     &       122     &       226     &       299     &       127     &       314     $\pm$   11      \\
108     &       399     &       305     &       147     &       227     &       289     &       146     &       297     $\pm$   12      \\
112     &       348     &       316     &       127     &       196     &       257     &       139     &       286     $\pm$   12      \\
117     &       391     &       321     &       139     &       120     &       267     &       232     &       294     $\pm$   22      \\
121     &       365     &       316     &       129     &       98      &       302     &       167     &       309     $\pm$   18      \\
125     &       319     &       291     &       154     &       105     &       314     &       185     &       303     $\pm$   20      \\
130     &       349     &       301     &       152     &       124     &       317     &       208     &       309     $\pm$   20      \\
134     &       376     &       315     &       155     &       124     &       326     &       211     &       321     $\pm$   21      \\
139     &       284     &       300     &       155     &       112     &       336     &       192     &       318     $\pm$   20      \\
143     &       185     &       308     &       173     &       104     &       369     &       222     &       338     $\pm$   25      \\
147     &       129     &       327     &       205     &       87      &       356     &       183     &       342     $\pm$   27      \\
152     &       99      &       304     &       159     &       59      &       321     &       145     &       312     $\pm$   25      \\
156     &       116     &       307     &       174     &       19      &       455     &       324     &       381     $\pm$   76      \\
160     &       116     &       312     &       191     &       8       &       432     &       214     &       372     $\pm$   78      \\
165     &       84      &       333     &       176     &       7       &       396     &       180     &       364     $\pm$   71      \\
169     &       76      &       335     &       176     &               &               &               &       335     $\pm$   20      \\
174     &       48      &       339     &       206     &               &               &               &       339     $\pm$   30      \\
178     &       22      &       376     &       194     &               &               &               &       376     $\pm$   41      \\
187     &       17      &       350     &       254     &               &               &               &       350     $\pm$   62      \\
191     &       21      &       348     &       221     &               &               &               &       348     $\pm$   48      \\
196     &       16      &       453     &       226     &               &               &               &       453     $\pm$   56      \\
200     &       20      &       388     &       207     &               &               &               &       388     $\pm$   46      \\
204     &       15      &       382     &       131     &               &               &               &       382     $\pm$   34      \\
208     &       3       &       304     &       82      &               &               &               &       304     $\pm$   48      \\
\end{supertabular}                                        

\vfill\eject

\tablecaption{Optical rotation curve of NGC 5985 at 4\arcsec\, resolution.  
\label{tab:rc_5985_ha}}
\tablehead{
\multicolumn{6}{l}{{\bf Table~\ref{tab:rc_5985_ha}} {\sl continued}}\\
\noalign{\smallskip}
\hline
\noalign{\smallskip}
R & N$_{app}$ & V$_{app}$ & $\sigma_{ring}$ & N$_{rec}$ & V$_{rec}$ & $\sigma_{ring}$ & V \\  
arcsec && \kms & \kms & & \kms & \kms & \kms  \\
\noalign{\smallskip}
\hline
\noalign{\smallskip}
}
\tablefirsthead{
\hline
\noalign{\smallskip}
R & N$_{app}$ & V$_{app}$ & $\sigma_{ring}$ & N$_{rec}$ & V$_{rec}$ & $\sigma_{ring}$ & V \\  
arcsec && \kms & \kms & & \kms & \kms & \kms  \\
\noalign{\smallskip}
\hline
\noalign{\smallskip}
}
\tabletail{
\noalign{\smallskip}
\hline
}
\begin{supertabular}{llllllll}  
11.5    & 1     & 146 &         -       &       &        &              & 146 $\pm$     -       \\
14.3    & 6     & 197 &         20      & 10    & 121    &      179     & 150 $\pm$     45      \\
18.4    & 11    & 223 &         34      & 32    & 222    &      37      & 223 $\pm$     8       \\
22.0    & 29    & 233 &         44      & 43    & 228    &      27      & 230 $\pm$     6       \\
25.8    & 73    & 250 &         38      & 43    & 246    &      31      & 248 $\pm$     5       \\
29.8    & 82    & 265 &         28      & 44    & 244    &      29      & 258 $\pm$     4       \\
33.6    & 89    & 270 &         17      & 38    & 247    &      36      & 263 $\pm$     4       \\
37.5    & 89    & 274 &         24      & 33    & 267    &      60      & 272 $\pm$     6       \\
41.7    & 105   & 287 &         22      & 36    & 286    &      42      & 287 $\pm$     4       \\
45.7    & 125   & 293 &         16      & 29    & 272    &      57      & 289 $\pm$     5       \\
49.4    & 134   & 298 &         27      & 45    & 273    &      25      & 292 $\pm$     3       \\
53.4    & 147   & 312 &         31      & 45    & 272    &      25      & 302 $\pm$     3       \\
57.6    & 155   & 313 &         49      & 60    & 275    &      24      & 302 $\pm$     4       \\
61.4    & 169   & 319 &         50      & 104   & 279    &      29      & 304 $\pm$     3       \\
65.2    & 141   & 318 &         33      & 88    & 278    &      38      & 303 $\pm$     3       \\
69.3    & 130   & 320 &         32      & 89    & 283    &      46      & 305 $\pm$     4       \\
73.4    & 112   & 315 &         30      & 93    & 282    &      40      & 300 $\pm$     3       \\
77.3    & 125   & 313 &         39      & 117   & 276    &      31      & 295 $\pm$     3       \\
81.1    & 108   & 301 &         45      & 109   & 287    &      28      & 294 $\pm$     4       \\
85.2    & 82    & 298 &         50      & 103   & 284    &      33      & 290 $\pm$     4       \\
89.2    & 86    & 292 &         45      & 113   & 290    &      35      & 291 $\pm$     4       \\
93.1    & 102   & 283 &         37      & 116   & 288    &      31      & 286 $\pm$     3       \\
97.1    & 141   & 284 &         34      & 92    & 292    &      37      & 287 $\pm$     3       \\
101.0   & 169   & 292 &         27      & 63    & 298    &      39      & 294 $\pm$     3       \\
105.0   & 160   & 288 &         29      & 55    & 297    &      34      & 290 $\pm$     3       \\
108.8   & 153   & 289 &         33      & 64    & 287    &      30      & 288 $\pm$     3       \\
112.7   & 124   & 295 &         30      & 45    & 275    &      43      & 290 $\pm$     4       \\
116.9   & 128   & 296 &         29      & 36    & 291    &      46      & 294 $\pm$     4       \\
120.7   & 108   & 298 &         22      & 26    & 294    &      43      & 297 $\pm$     4       \\
124.8   & 91    & 296 &         22      & 29    & 283    &      33      & 293 $\pm$     4       \\
128.6   & 93    & 294 &         36      & 14    & 275    &      25      & 291 $\pm$     4       \\
132.8   & 93    & 304 &         18      & 19    & 295    &      31      & 303 $\pm$     3       \\
136.6   & 85    & 309 &         24      & 20    & 289    &      30      & 305 $\pm$     4       \\
140.5   & 76    & 322 &         37      & 15    & 284    &      18      & 316 $\pm$     4       \\
144.4   & 21    & 296 &         52      & 10    & 282    &      13      & 291 $\pm$     10      \\
147.9   & 9     & 276 &         36      & 6     & 272    &      10      & 275 $\pm$     10      \\
76.3    & 18    & 290 &         43      &       &        &              & 290 $\pm$     10      \\
78.2    & 21    & 314 &         14      &       &        &              & 314 $\pm$     3       \\
79.9    & 17    & 296 &         64      &       &        &              & 296 $\pm$     16      \\
82.2    & 30    & 310 &         13      &       &        &              & 310 $\pm$     2       \\
164.4   & 19    & 303 &         17      &       &        &              & 303 $\pm$     4       \\
168.0   & 4     & 275 &         10      &       &        &              & 275 $\pm$     5       \\
\end{supertabular}
\vfill\eject

\tablecaption{\hI rotation curve of NGC 5985 at $\sim$10\arcsec\, binning (30\arcsec\, resolution).
\label{tab:rc_5985_hI}} 
\tablehead{ 
\multicolumn{6}{l}{{\bf Table~\ref{tab:rc_5985_hI}} {\sl continued}}\\ 
\noalign{\smallskip} 
\hline 
\noalign{\smallskip} 
R & V \\ 
arcsec & \kms \\ 
\noalign{\smallskip} 
\hline 
\noalign{\smallskip} 
} 
\tablefirsthead{ 
\hline \hline 
\noalign{\smallskip} 
R & V \\ 
arcsec & \kms \\ 
\noalign{\smallskip} 
\hline 
\noalign{\smallskip} 
} 
\tabletail{ 
\noalign{\smallskip} 
\hline 
} 
\begin{supertabular}{ll} 
  60  & 294.2    $\pm$  9.21 \\
  75  & 299.0    $\pm$  5.94 \\ 
  90  & 297.3    $\pm$  3.31 \\ 
 105  & 295.6    $\pm$  3.42 \\ 
 120  & 295.9    $\pm$  3.23 \\ 
 135  & 297.1    $\pm$  3.72 \\ 
 150  & 292.4    $\pm$  4.49 \\ 
 165  & 288.6    $\pm$  5.40 \\ 
 180  & 285.1    $\pm$  5.89 \\ 
\end{supertabular} 
}
\vfill\eject  

\end{document}